\documentclass[%
reprint,
superscriptaddress,
onecolumn,
nofootinbib,
amsmath,amssymb,
aps,
prd,
]{revtex4-2}

\usepackage{graphicx}                   
\usepackage{dcolumn}                    
\usepackage{bm}                         
\usepackage{hyperref}                   
\usepackage{physics}                   
\usepackage[noabbrev]{cleveref}         
\usepackage{braket}                     
\usepackage{mathtools}  
\usepackage{tikz}
\newcommand{\rev}[1]{{#1}}

\newcommand{\Sp}[1]{\ensuremath{\mathrm{Sp}(#1)}}
\newcommand{\SU}[1]{\ensuremath{\mathrm{SU}(#1)}}
\newcommand{\SO}[1]{\ensuremath{\mathrm{SO}(#1)}}
\newcommand{\U}[1]{\ensuremath{\mathrm{U}(#1)}}
\graphicspath{{figures/}}

\newcommand{\beq}{\begin{equation}}
\newcommand{\eeq}{\end{equation}}
\newcommand{\beqs}{\begin{eqnarray}}
\newcommand{\eeqs}{\end{eqnarray}}

\begin{document}
\title{Singlets in gauge theories with fundamental matter}

\author{Ed Bennett}
\email{e.j.bennett@swansea.ac.uk}
\affiliation{Swansea Academy of Advanced Computing, Swansea University (Bay Campus), Fabian Way, SA1 8EN Swansea, Wales, United Kingdom}
\author{Ho Hsiao}
\email{thepaulxiao.sc06@nycu.edu.tw}
\affiliation{Institute of Physics, National Yang Ming Chiao Tung University, 1001 Ta-Hsueh Road, Hsinchu 30010, Taiwan}
\author{Jong-Wan Lee}
\email{j.w.lee@ibs.re.kr}
\affiliation{
Particle Theory  and Cosmology Group, Center for Theoretical Physics of the Universe, Institute for Basic Science (IBS),  Daejeon, 34126, Korea}
\affiliation{Department of Physics, Pusan National University, Busan 46241, Korea}
\affiliation{Institute for Extreme Physics, Pusan National University, Busan 46241, Korea}
\author{Biagio Lucini}
\email{b.lucini@swansea.ac.uk}
\affiliation{Swansea Academy of Advanced Computing, Swansea University (Bay Campus), Fabian Way, SA1 8EN Swansea, Wales, United Kingdom}
\affiliation{Department of Mathematics, Faculty of Science and Engineering, Swansea University (Bay Campus), Fabian Way, SA1 8EN Swansea, Wales, United Kingdom}
\author{Axel Maas}
\email{axel.maas@uni-graz.at}
\affiliation{Institute of Physics, NAWI Graz, University of Graz, Universit\"atsplatz 5, A-8010 Graz, Austria}
\author{Maurizio Piai}
\email{m.piai@swansea.ac.uk}
\affiliation{Department of Physics, Faculty of Science and Engineering, Swansea University (Park Campus), Singleton Park, SA2 8PP Swansea, Wales, United Kingdom}
\author{Fabian Zierler}
\email{fabian.zierler@uni-graz.at}
\affiliation{Institute of Physics, NAWI Graz, University of Graz, Universit\"atsplatz 5, A-8010 Graz, Austria}

\date{\today}

\begin{abstract}

We provide the first determination of the mass of the lightest flavor-singlet pseudoscalar and scalar bound states (mesons), in the \Sp{4} Yang-Mills theory coupled to two flavors of fundamental fermions, using lattice methods.  This theory has applications both to composite Higgs and strongly-interacting dark matter scenarios. We find the singlets to have masses comparable to those of the light flavored states, which might have important implications for phenomenological models.  We focus on  regions of parameter space corresponding to a moderately heavy mass regime for the fermions. We compare the spectra we computed to existing and new results for \SU{2} and \SU{3} theories, uncovering an intriguing degree of commonality. As a by-product,  in order to perform the aforementioned measurements, we implemented and tested, in the context of symplectic lattice gauge theories, several strategies for the treatment of disconnected-diagram contributions to two-point correlation functions. These technical advances set the stage for future  studies of the singlet sector in broader portions of parameter space of this and other lattice theories with a symplectic gauge group.

\end{abstract}
\maketitle

\tableofcontents

\section{Introduction}

The possible existence of new strongly interacting sectors that extend  the standard model (SM) of particle physics has been the subject of a long-standing history of theoretical studies. In recent years, this idea has prominently featured in the context of composite Higgs models (CHMs) in which the Higgs fields of the SM  originate as pseudo-Nambu-Goldstone bosons (PNGBs) of the underlying theory~\cite{Kaplan:1983fs,Georgi:1984af,Dugan:1984hq}.\footnote{The recent literature is vast. See, e.g., the reviews in Refs.~\cite{Panico:2015jxa,Witzel:2019jbe, Cacciapaglia:2020kgq}, the summary tables in Refs.~\cite{Ferretti:2013kya,Ferretti:2016upr,Cacciapaglia:2019bqz},and the selection of papers in Refs.~\cite{Katz:2005au,Barbieri:2007bh,Lodone:2008yy,Gripaios:2009pe,Mrazek:2011iu,Marzocca:2012zn,Barnard:2013zea,Grojean:2013qca,Cacciapaglia:2014uja,Ferretti:2014qta,Arbey:2015exa,Cacciapaglia:2015eqa,Feruglio:2016zvt,DeGrand:2016pgq,Fichet:2016xvs,Galloway:2016fuo,Agugliaro:2016clv,Belyaev:2016ftv,Csaki:2017cep,Chala:2017sjk,Golterman:2017vdj,Csaki:2017jby,Alanne:2017rrs,Alanne:2017ymh,Sannino:2017utc,Alanne:2018wtp,Bizot:2018tds,Cai:2018tet,Agugliaro:2018vsu,Cacciapaglia:2018avr,Gertov:2019yqo,Ayyar:2019exp,Cacciapaglia:2019ixa,BuarqueFranzosi:2019eee,Cacciapaglia:2019dsq,Cacciapaglia:2020vyf,Dong:2020eqy,Cacciapaglia:2021uqh,Banerjee:2022izw} and Refs.~\cite{Contino:2003ve,Agashe:2004rs,Agashe:2005dk,Agashe:2006at,Contino:2006qr,Falkowski:2008fz,Contino:2010rs,Contino:2011np,Caracciolo:2012je,Erdmenger:2020lvq,Erdmenger:2020flu,Elander:2020nyd,Elander:2021bmt,Elander:2021kxk,Elander:2023aow}.} 
A parallel development has led to strongly coupled gauge theories being considered as the  dynamical origin of hidden sectors in which dark matter consists of strongly interacting massive particles (SIMPs)~\cite{Hochberg:2014dra,Hochberg:2014kqa,Hochberg:2015vrg,Hansen:2015yaa, Bernal:2017mqb,Berlin:2018tvf,Bernal:2019uqr,Tsai:2020vpi,Kondo:2022lgg,Bernal:2015xba}. This scenario  
can address observational problems such as the  {\it `core vs. cusp'}~\cite{deBlok:2009sp} and {\it `too big to fail'}~\cite{Boylan-Kolchin:2011qkt} ones, and 
have implications for  gravitational wave experiments~\cite{Seto:2001qf,
Kawamura:2006up,Crowder:2005nr,Corbin:2005ny,Harry:2006fi,
Hild:2010id,Yagi:2011wg,Sathyaprakash:2012jk,Thrane:2013oya,
Caprini:2015zlo,
LISA:2017pwj,
LIGOScientific:2016wof,Isoyama:2018rjb,Baker:2019nia,
Brdar:2018num,Reitze:2019iox,Caprini:2019egz,
Maggiore:2019uih}---see e.g.\
Refs.~\cite{Witten:1984rs,Kamionkowski:1993fg,Allen:1996vm,Schwaller:2015tja, Croon:2018erz,Christensen:2018iqi},
as well as Refs.~\cite{Huang:2020crf,Halverson:2020xpg,Kang:2021epo,Reichert:2021cvs,Reichert:2022naa}.
Both CHMs and SIMPs give rise to particles, the PNGBs, carrying non-trivial quantum numbers of non-Abelian global (flavor) symmetries, that suppress and protect their masses.

In general, strongly coupled theories also yield bound states that are flavor singlets.
 Composite models with a strongly coupled origin can in particular give rise to a light dilaton~\cite{Leung:1985sn,Bardeen:1985sm,Yamawaki:1985zg},  the PNGB associated with (approximate) scale invariance. The striking phenomenological implications of this possibility~\cite{Goldberger:2007zk} have also been studied extensively (see e.g.\  Refs.~\cite{Hong:2004td,Dietrich:2005jn,Hashimoto:2010nw,Appelquist:2010gy,Vecchi:2010gj,Chacko:2012sy,Bellazzini:2012vz,Bellazzini:2013fga,Abe:2012eu,Eichten:2012qb,Hernandez-Leon:2017kea,CruzRojas:2023jhw}),  and the low energy effective field theory (EFT) of the dilaton~\cite{Migdal:1982jp,Coleman:1985rnk}
and the other PNGBs can be combined in the dilaton EFT~\cite{Matsuzaki:2013eva,Golterman:2016lsd,Kasai:2016ifi,Hansen:2016fri,Golterman:2016cdd,Appelquist:2017wcg,Appelquist:2017vyy,Golterman:2018mfm,Cata:2019edh,Cata:2018wzl,Appelquist:2019lgk,Golterman:2020tdq,Golterman:2020utm,Appelquist:2020bqj,Appelquist:2022qgl,Appelquist:2022mjb}. Numerical evidence supporting this possibility 
has emerged in the context of $\SU{3}$ lattice gauge theories with special field content~\cite{LatKMI:2014xoh,Appelquist:2016viq,LatKMI:2016xxi,Gasbarro:2017fmi,LatticeStrongDynamics:2018hun,LatticeStrongDynamicsLSD:2021gmp,Hasenfratz:2022qan,Fodor:2012ty,Fodor:2015vwa, Fodor:2016pls,Fodor:2017nlp,Fodor:2019vmw,Fodor:2020niv}.
The singlet sector of a strongly coupled theory  contains also pseudoscalar composite states, 
the phenomenology of which is the subject of  dedicated studies~\cite{Belyaev:2016ftv,Cai:2015bss,Belyaev:2015hgo,DeGrand:2016pgq,Cacciapaglia:2017iws,Cacciapaglia:2019bqz}. Their EFT treatment  follows closely that of axion-like particles (ALPs)~\cite{Brivio:2017ije,Bellazzini:2017neg,Bauer:2017ris}.

The  phenomenological consequences of such theories depend crucially on the mass spectrum of the lightest states, and on the (model dependent) couplings to the standard model. While an efficient tool in treating the latter is provided by the EFT methodology, even the construction of such EFTs requires a good understanding of the lightest portion of the spectrum. 
Whatever the original motivation and application envisioned for such new strongly coupled physical sectors is, they have a plethora of bound states, some of which can be either stable or long-lived, and potentially light. It is then desirable to gain a broad non-perturbative understanding of their spectroscopy, for all bound states,
and in the largest possible portions of the parameter space.
The instrument of choice for this endeavor is that of (numerical) lattice gauge theories. There has been a wide variety of investigations into the spectrum of many such theories, especially  those with  \SU{N} group, with matter in various numbers and representations. Besides the aforementioned work,
and Refs.~\cite{Aoki:2017fnr, LatKMI:2015rtu, Brower:2015owo, Fodor:2017rro, LatticeStrongDynamics:2018hun,LatticeStrongDynamicsLSD:2021gmp, LatticeStrongDynamicsLSD:2014osp}  on $\SU{3}$, see e.g.\ Refs.~\cite{Drach:2021uhl, Drach:2020wux, Drach:2017btk, Hietanen:2014xca, Detmold:2014kba, Lewis:2011zb, Arthur:2016dir, Francis:2018xjd,Drach:2015epq} for $\SU{2}$ theories, Refs.~\cite{DeGrand:2015lna, Ayyar:2017qdf, Ayyar:2018zuk, Cossu:2019hse} for $\SU{4}$ in multiple representations, and \cite{Wellegehausen:2013cya} for $\mathrm{G}_2$, as well as the reviews in Refs.~\cite{DeGrand:2015zxa, DeGrand:2019vbx, Drach:2020qpj, Rummukainen:2022ekh} 

Gauge theories with symplectic group play a special role in all these contexts, because of the peculiar properties of \Sp{2N} groups and their representations. For example, the model in Ref.~\cite{Barnard:2013zea}
provides the simplest realization of a CHM that combines it with top partial compositeness~\cite{Kaplan:1991dc}, and consists of a  $\Sp{2N}$ gauge theory with mixed-representation fermion content. Likewise, Refs.~\cite{Hochberg:2014dra,Hochberg:2015vrg} use it for the construction of a SIMP-``miracle". A number of recent lattice studies started to characterize them~\cite{Bennett:2017kga,
Lee:2018ztv,Bennett:2019jzz,
Bennett:2019cxd,Bennett:2020hqd,
Bennett:2020qtj,Lucini:2021xke,
Bennett:2021mbw,Bennett:2022yfa,
Bennett:2022gdz,Bennett:2022ftz,
Hsiao:2022gju,Hsiao:2022kxf,
Maas:2021gbf,Zierler:2021cfa,
Kulkarni:2022bvh,Bennett:2023wjw}, following the pioneering effort in Ref.~\cite{Holland:2003kg}.
While much such work has focused on the spectroscopy of bound states carrying flavor,
 with this paper we report on progress in the singlet sector.

The present study explores the flavor-singlet bound state sector in the  \Sp{4} gauge theory coupled to two fermions in the fundamental representation,
a theory 
that has gathered substantial interest in the CHM context,
and also provides a minimal realization of the SIMP mechanism~\cite{Hansen:2015yaa, Berlin:2018tvf, Beylin:2020bsz, Tsai:2020vpi, Kondo:2022lgg}. 
This study is complementary to the available non-singlet hadron spectrum found in Refs.~\cite{Bennett:2022yfa, Kulkarni:2022bvh, Bennett:2021mbw, Bennett:2020qtj, Bennett:2020hqd, Bennett:2019jzz, Bennett:2019cxd, Bennett:2017kga}.
Within the general aim of understanding universal features of the low-lying spectrum across different gauge groups, this is also a step towards understanding  how the approach to the large-$N$ limit of \Sp{2N} gauge theories differs from that of \SU{N} gauge theories, especially with respect to the axial anomaly and topology. Our results could in the future help  to understand the anomaly-mediated decays of singlets into SM particles. 

We supplement this publication with numerical results obtained in two other theories.
The first is the closely related  \SU{2} theory with two fundamental fermions, for which earlier  studies exist~\cite{Drach:2021uhl, Arthur:2016ozw}, and for which we perform additional, new calculations.
In the case of the  \SU{3} theory coupled to fermions in the fundamental representation, the pseudoscalar singlet  has been studied in Refs.~\cite{UKQCD:1998zbe,McNeile:2000hf,TXL:2000mhy,Bali:2000vr,CP-PACS:2002exu,Allton:2004qq,Urbach:2007rt,Jansen:2008wv,Hashimoto:2008xg,Sun:2017ipk,Dimopoulos:2018xkm,Fischer:2020yvw,Jiang:2022ffl}. The determination of the mass (and width) of the lightest scalar singlet  has proven to be particularly challenging and a number of studies in \SU{3} with dynamical fermions exists both in the context of real-world QCD~\cite{Kunihiro:2003yj, Hart:2006ps,  Prelovsek:2010kg, Fu:2012gf, Dudek:2014qha, Wakayama:2014gpa, Briceno:2016mjc, Briceno:2017qmb, Guo:2018zss, Rodas:2023gma} and more general field content~\cite{Aoki:2017fnr, LatKMI:2015rtu, LatKMI:2016xxi, Brower:2015owo, Fodor:2017rro, LatticeStrongDynamics:2018hun,LatticeStrongDynamicsLSD:2021gmp}.
We borrow results from this extensive literature, for the purpose of comparing with our own results.

The paper is organized as follows.
The pseudo-real nature of the fundamental representation of \Sp{4}---as for \SU{2}---leads to symmetry enhancement by modifying the flavor symmetry and symmetry-breaking pattern. The structure of the low-lying spectrum is hence different from the more familiar QCD case. We briefly comment on the most striking such features  in Sect.~\ref{sec:structure}, as we define the continuum theory of interest. In Sect.~\ref{sec:lattice} we describe the lattice methods that we use to study the flavor-singlet states, putting some emphasis on the implications for the construction of suitable operators,  in Sect.~\ref{ssec:op}. The study of correlations functions involving  singlets is affected by  notorious difficulties, poor signal-to-noise ratio featuring prominently among them. This required the adoption of advanced techniques to obtain a non-zero signal, as we explain  in Sect.~\ref{ssec:signalboost}, and in more detail in the Appendices. 

We present the body of our numerical results in Sect.~\ref{sec:results}. Section~\ref{ssec:psdeg} is devoted to the lightest pseudoscalar singlet state, in both the \Sp{4} and \SU{2} theories, for degenerate masses. We report on  the case of non-degenerate flavor masses for \Sp{4} (which realizes a scenario relevant to dark matter models) in 
Sect.~\ref{ssec:psnondeg}. The scalar singlet sector,  in the degenerate case for the  \Sp{4} theory, is the subject of Sect.~\ref{ssec:sdeg}, though we anticipate that, because of the bad signal-to-noise ratio, only a rough estimate with unclear finite-spacing systematics can be established at this stage. Finally, we assess our  results by a comparison to the \SU{3} case, and report the results in Sect.~\ref{ssec:compsu3}. Our general conclusion, exposed more critically in Sect.~\ref{sec:summary}, is that for the available range of fermion masses the singlets are indeed light enough to affect phenomenology and low-energy EFT considerations. 
We add several technical appendices, covering further details. We note that some preliminary 
results are available in Ref.~\cite{Zierler:2022qfq}.

\section{Flavor singlets in symplectic gauge theories}
\label{sec:structure}

We start the presentation by defining explicitly the (continuum) field theories of interest. We provide both their
microscopic definition, in terms of elementary fields, and their salient long-distance properties, which
can be explained  in EFT terms.  In doing so, we emphasize the role of the symmetries of the theory.

\subsection{Microscopic theory and global symmetries}

The $\Sp{2N}$ gauge theories of interest are characterized by a Lagrangian density, ${\cal L}$, which in
this section we write using a metric with Lorentzian signature $(+1,-1,-1,-1)$, and takes the form:
\beqs
\label{Eq:L}
    \mathcal L = -\frac{1}{2} \Tr \left[G_{\mu \nu}G^{\mu \nu}\right] 
    + \bar u \left(i \gamma^\mu D_\mu - m_u\right) u + \bar d \left( i\gamma^\mu D_\mu - m_d\right) d\,,
\eeqs
where $G_{\mu\nu}\equiv \sum_A G_{\mu\nu}^A t^A$ are the field strength tensors, and $t^A$ are the generators of $\Sp{2N}$, normalized so that $\Tr T^AT^B=\frac{1}{2}\delta^{AB}$, while $u$ and $d$ are 4-component Dirac spinors, denoting the two flavors of fermion fields transforming in the fundamental representation of $\Sp{2N}$.
The Lagrangian is real and Lorentz invariant, as $\bar u \equiv u^{\dagger} \gamma^0$, and 
$\bar d \equiv d^{\dagger} \gamma^0$. The covariant derivatives are
defined in terms of the gauge fields $A_{\mu}\equiv A_{\mu}^A t^A$ as
\beqs
D_{\mu} u &\equiv& \partial_{\mu} u + i g A_{\mu} u\,,~~~~~~~~~
D_{\mu} d \,\equiv\, \partial_{\mu} d + i g A_{\mu} d\,,
\eeqs
where $g$ is the gauge coupling. Explicitly, the field-strength tensor is given by
\beqs
G_{\mu \nu} &\equiv& \partial_{\mu} A_{\nu} -\partial_{\nu} A_{\mu} +i g \left[A_{\mu} \,,\,A_{\nu}\right]\,,
\eeqs
where $[\cdot,\cdot]$ denotes the commutator.

The fundamental representation of $\Sp{2N}$ is pseudo-real. As a result, the global symmetry is enhanced: one can show explicitly, by rewriting Eq.~(\ref{Eq:L}) in terms of 2-component fermions,\footnote{This somewhat tedious exercise can be found in all its details in the literature, for instance in Refs.~\cite{Lewis:2011zb,Bennett:2019cxd} and references therein.} that for each Dirac fermion the $\U{1}_L\times \U{1}_R$ Abelian global symmetry acting on its chiral projections is enhanced to a non-Abelian $\U{2}$ global symmetry. The fermion kinetic terms hence, written in terms of covariant derivatives, exhibit an enhanced (classical) $\U{4}=U(1)_A\times SU(4)$ global symmetry; we will return to the effect of anomalies in the next subsection.

The fermion mass terms break explicitly the global symmetry. 
If the masses are degenerate, $m_u=m_d$, as will be the case throughout most of this paper, then the global symmetry is explicitly  broken to $\Sp{4}$. The bilinear, non-derivative operator appearing in the Lagrangian density as a mass term  is also expected to condense, so that the same symmetry breaking pattern appears also in
 spontaneous symmetry breaking effects.
 For generic choices of fermion masses, $m_u\neq m_d$, the approximate global symmetry is further broken to $\Sp{2}^2\sim\SU{2}^2$ \cite{Kulkarni:2022bvh}. 
 
\subsection{Light meson spectrum  for two fundamental fermions}
\label{sec:degenerate}

We summarize here the main properties of the bound states of interest,
guided by gauge invariance and symmetry arguments, starting from the case in which the fundamental fermions have degenerate mass, $m_u=m_d$. We observe that the group structure has an even number of colors,
hence  baryons are bosons. Furthermore,
because the matter content consists only of fermions transforming in the pseudo-real fundamental representation, and
as a result  the global symmetry is enhanced, ordinary baryon number  is a subgroup of the enhanced $\SU{4}$, and is unbroken in the vacuum of the theory, 
and hence objects that one might be tempted to classify as having different baryon number
(e.g.\ mesons and diquarks) belong to the same \Sp{4} multiplet. 

We restrict the discussion from here onwards to mesons made of two fundamental fermions.
It may be convenient to think of the mesons in terms of their 2-component fermion field content, in order to  classify them 
by their $\Sp{4}$ transformations, and attribute their $J^P$ quantum numbers.\footnote{We define the parity $P$ so that flavor eigenstates are also parity eigenstates. See Refs.~\cite{Drach:2017btk,Bennett:2019cxd,Kulkarni:2022bvh} for extended discussions of the subtleties involving twisting between space and flavor  in the definition of parity.}
As the 2-component fermions transform as a $4$ of $\Sp{4}$, the multiplication properties imply that there exist 
mesons transforming as a $1$, $5$, and $10$ of \Sp{4}.

Starting from the spin-0 states, one expects to find in the spectrum $5$ PNGBs, spanning the $\SU{4}/\Sp{4}$ coset, and transforming as a $5$ of $\Sp{4}$,  to become massless in the $m_u=m_d\rightarrow 0$ limit.\footnote{These are the states playing a role in CHMs; in their EFT description in terms of weakly coupled fields, four of them are identified with the components of the SM Higgs doublet, because the $\SU{2}_L\times \SU{2}_R$ approximate symmetry of the electroweak theory is identified with the $SO(4)$ subgroup of \Sp{4}, and hence $5=4\oplus 1$~\cite{Barnard:2013zea}. The $1$ is an additional state, that carries no SM gauge quantum numbers, but in the EFT description of the strong coupling sector it is degenerate with the $4$. Likewise, these states are the dark matter candidates in SIMP models \cite{Hochberg:2014dra,Hochberg:2014kqa,Kulkarni:2022bvh}.}
These states have parity partners, generalizing what in QCD literature are usually denoted as $a_0$ particles. Some numerical lattice evidence exists that at high temperature these two sets of states become degenerate---see Ref.~\cite{Lee:2017uvl} for $\SU{2}$ and Ref.~\cite{Brandt:2016daq} for $\SU{3}$, both with two flavors of fundamental fermions---because the $\U{1}_A$ symmetry relating them is restored.
But at zero temperature the scalar $5$ is expected to be heavy, the mass of the particles being of the order of the confinement scale, even in the  $m_u=m_d\rightarrow 0$ limit. 

Classically, one would expect also two singlet scalars to be light: the axion and the dilaton. Indeed, the classical Lagrangian for $m_u=m_d= 0$ is invariant also under the action of a $\U{1}_A$ symmetry and of dilatations, the condensates breaking both of them spontaneously, and these two additional light states can be thought of as the PNGBs  associated with these two Abelian symmetries.
Alas, besides being explicitly broken by the fermion masses,  
both these symmetries are also anomalous. 
The  $\U{1}_A$ and scale anomalies hence provide  masses for the axion-dilaton system, related to the scale of confinement of the theory.  Mixing effects between these states and other vacuum excitations (e.g.~glueballs with the same $J^P$ quantum numbers) are present as well, given that no symmetry argument can be invoked rigorously to forbid them. The precise determination of such masses, hence, is non-trivial, 
and to large extent this paper is about setting the stage for its future large-scale, high-precision calculation.
Furthermore, in confining theories with large numbers of degrees of freedom, and when approaching
the lower edge of the conformal window,  non-perturbative effects might suppress the mass of the axion and dilaton, respectively; this is a very active field of research in itself, for a potential phenomenological role both of this axion
and of the dilaton, as we mentioned in the introduction;
the technology we  developed and tested for this paper
could play an important future role in either case.

The spin-1 part of the meson spectrum is more rich. In analogy with the case of QCD, one expects the lightest such
states to generalize the $\rho$ mesons; they transform as a $10$ of $\Sp{4}$ and have $J^P=1^-$, and their properties have been studied elsewhere~\cite{Bennett:2019jzz}. They have the peculiar property that they can be sourced by two  different interpolating operators, with the schematic structures $\bar{\psi} \gamma_{\mu} \psi$  and 
$\bar{\psi} \sigma_{\mu\nu} \psi$, respectively. In addition, the generalizations of the $a_1$ and $b_1$ from QCD
transform as a $5$ and a $10$ of $\Sp{4}$, respectively; these additional states are heavier than  the aforementioned $10$-plet with  $J^P=1^-$---see the discussions in Refs.~\cite{Lewis:2011zb,Bennett:2019cxd}. It is worth noticing that some spin-1 singlet mesons of QCD are actually part of these multiplets, because of the symmetry-enhancement pattern---noticeably, the particle that corresponds to $\omega$ in QCD.

In the presence of non-degenerate fermion masses, $m_u\neq m_d$,
 the global symmetry breaks further from $\Sp{4}$ down to $\Sp{2}\times \Sp{2}=\SU{2} \times \SU{2}\sim \SO{4}$. Consequently, the multiplets decompose with respect to the smaller flavor symmetry \cite{Kulkarni:2022bvh}. 
 The $5$-plets split into a $4$-plet  and a singlet, whereas the $10$-plet decomposes into a $6$-plet and a $4$-plet.
 This implies that an additional singlet appears in states that would have been a $5$-plet in the mass-degenerate theory, such as the PNGBs and the axial-vectors. This is the familiar scenario in QCD: in the presence of a mass difference between up and down quark,  isospin is explicitly broken and the flavor-neutral pion $\pi^0$ becomes a singlet, with different mass from the charged $\pi^{\pm}$ states. The main difference with QCD is that, as the pseudo-reality of the representation results in additional flavor neutral states, which microscopically can be written as di-quark states, the multiplet is enlarged.
 
This  differs for mesons in the $10$-plet representation, such as the vector meson. The $10$ decomposes in a $4$-plet (with the same flavor structure as in the case of the PNGBs) and a $6$-plet. The latter consists of two states sourced by ordinary meson operators (in the QCD analogy, they are  the $\rho^0$ and the $\omega$  particles with $J^P=1^-$) and four other states that are sourced by diquark operators. 
A further splitting of these multiplets is possible, for example by gauging a 
 $\U{1}$ subgroup of the $\SO{4}$ symmetry~\cite{Kulkarni:2022bvh}, but we do not consider it here.

\section{Lattice setup}\label{sec:lattice}
We perform lattice simulations using the standard plaquette action and standard Wilson fermions \cite{Wilson:1974sk}. We use the HiRep code \cite{DelDebbio:2008zf, HiRepSUN} extended for $\Sp{2N}$ gauge theories \cite{HiRepSpN} to generate configurations and to perform the measurements. In the case of degenerate fermions we use the Hybrid Monte Carlo (HMC) \cite{Duane:1987de} algorithm and for non-degenerate fermions we use the rational HMC (RHMC) \cite{Clark:2006fx} algorithm. The latter case does not guarantee positivity of the fermion determinant. In this case we have monitored the lowest eigenvalue of the Dirac operator which we always found to be positive. Thus, we do not see any hints of a sign problem for the fermion masses studied in this work. Results for the non-singlet spectrum for two fundamental fermions were first reported in Refs.\cite{Bennett:2019cxd,Kulkarni:2022bvh}. We give simulation details of our ensembles in Tabs.~\ref{tab:ensembles_degenerate} and~\ref{tab:ensembles_non-degenerate}.   

We perform simulations on Euclidean lattices of size $T \times L^3$ and define the bare inverse gauge coupling as $\beta = 8/g^2$. We implement periodic boundary conditions for the gauge fields. For the Dirac fields we impose periodic boundary conditions in the spatial directions and anti-periodic boundary conditions in the temporal direction.

\begin{table}
    \setlength{\tabcolsep}{3pt}
    \begin{tabular}{|c|c|c|c|c|c|c|c|c|c|c|c|c|c|}
	\hline
	Ensemble & group & $\beta$ & $m_0$ & $L$ & $T$ & $n_\text{conf}$ & $n_\text{src}$ & $I_{\eta'}$ & $I_\pi$ & $I_\sigma$ & $I_{\sigma^{\rm conn.}}$ & $I_\rho$ &  $\langle P \rangle$\\
	\hline\hline
	SU2B1L1M8&SU(2)&2.0&-0.947&20&32&1020&300&-&(10, 16)&(5, 8)&(7, 10)&(10, 16)&0.56734(2)\\
	SU2B1L1M7&SU(2)&2.0&-0.94&14&24&1851&192&(8, 12)&(8, 12)&-&-&(9, 12)&0.56516(3)\\
	SU2B1L1M6&SU(2)&2.0&-0.935&16&32&951&256&(7, 11)&(9, 16)&-&-&(9, 16)&0.563654(28)\\
	SU2B1L1M5&SU(2)&2.0&-0.93&14&24&1481&256&(7, 12)&(8, 12)&-&-&(9, 12)&0.56245(3)\\
	SU2B1L1M4&SU(2)&2.0&-0.925&14&24&1206&192&(6, 10)&(8, 12)&-&-&(9, 12)&0.56119(3)\\
	SU2B1L1M3&SU(2)&2.0&-0.92&12&24&2401&192&(6, 9)&(7, 12)&-&(6, 11)&(8, 12)&0.559983(29)\\
	SU2B1L1M2&SU(2)&2.0&-0.9&12&24&500&128&(6, 9)&(7, 12)&-&-&(8, 12)&0.55571(6)\\
	SU2B1L1M1&SU(2)&2.0&-0.88&10&20&2582&128&(5, 8)&(8, 10)&-&-&(9, 10)&0.55225(4)\\
	\hline\hline
	Sp4B3L1M8&Sp(4)&7.2&-0.799&32&40&451&224&-&(15, 20)&(5, 9)&(11, 19)&(15, 20)&0.590862(5)\\
	Sp4B3L1M7&Sp(4)&7.2&-0.794&28&36&504&288&(7, 12)&(10, 18)&-&(11, 16)&(11, 18)&0.590452(7)\\
	Sp4B3L1M6&Sp(4)&7.2&-0.79&24&36&500&320&(7, 12)&(12, 18)&(5, 8)&(10, 16)&(13, 18)&0.590127(9)\\
	Sp4B3L1M5&Sp(4)&7.2&-0.78&24&36&508&384&(6, 12)&(12, 18)&-&(11, 15)&(13, 18)&0.589278(8)\\
	Sp4B3L1M4&Sp(4)&7.2&-0.77&24&36&200&384&(6, 11)&(11, 18)&-&(10, 15)&(12, 18)&0.588460(12)\\
	Sp4B3L1M3&Sp(4)&7.2&-0.76&16&36&200&384&-&(11, 18)&(5, 8)&(9, 14)&(12, 18)&0.587666(25)\\
	\hline\hline
	Sp4B1L1M7&Sp(4)&6.9&-0.924&24&32&492&320&-&(9, 16)&(4, 7)&(7, 10)&(10, 16)&0.56317(2)\\
	Sp4B1L1M6&Sp(4)&6.9&-0.92&24&32&503&484&-&(7, 16)&(4, 9)&(8, 12)&(8, 16)&0.562075(13)\\
	Sp4B1L2M6&Sp(4)&6.9&-0.92&16&32&176&128&(6, 10)&(9, 16)&(4, 10)&(7, 10)&(9, 16)&0.56212(5)\\
	Sp4B1L1M5&Sp(4)&6.9&-0.91&16&32&435&256&(6, 11)&(8, 16)&-&(7, 9)&(9, 16)&0.55935(3)\\
	Sp4B1L2M5&Sp(4)&6.9&-0.91&14&24&513&256&(5, 10)&(8, 12)&(4, 7)&(9, 12)&(9, 12)&0.55941(3)\\
	Sp4B1L1M4&Sp(4)&6.9&-0.9&16&32&547&512&(6, 10)&(9, 16)&-&(7, 10)&(10, 16)&0.556921(25)\\
	Sp4B1L2M4&Sp(4)&6.9&-0.9&14&24&942&128&(7, 10)&(8, 12)&(4, 9)&(7, 9)&(9, 12)&0.556981(26)\\
	Sp4B1L3M4&Sp(4)&6.9&-0.9&12&24&2904&128&(6, 10)&(8, 12)&(4, 8)&(8, 10)&(9, 12)&0.557009(18)\\
	Sp4B1L2M3&Sp(4)&6.9&-0.89&14&24&461&128&(7, 10)&(8, 12)&(5, 9)&(8, 11)&(9, 12)&0.55468(4)\\
	Sp4B1L3M3&Sp(4)&6.9&-0.89&12&24&1019&320&(6, 10)&(8, 12)&(3, 6)&(7, 11)&(9, 12)&0.55467(3)\\
	Sp4B1L2M2&Sp(4)&6.9&-0.87&12&24&1457&128&(7, 11)&(8, 12)&(5, 8)&(8, 10)&(9, 12)&0.550497(27)\\
	Sp4B1L2M3&Sp(4)&6.9&-0.87&10&20&976&128&(6, 9)&(8, 10)&-&(6, 10)&(8, 10)&0.55068(5)\\
	\hline\hline
\end{tabular}
    \caption{List of all ensembles with degenerate fermion masses used in this work. We report the number of configurations $n_{\rm conf}$, the number of the stochastic sources used in the approximation of the all-to-all quark propagator $n_{\rm src}$, the intervals for fitting the resulting meson correlators $I_{\rm meson}$ and the average value of the plaquette $\langle P \rangle$. In some cases we were unable to identify a clear plateau in the effective masses and could not determine the singlet masses. In these cases we do not report a fit interval. For the singlet mesons the interval quoted here was used to fit the correlators after subtracting the excited state contributions in the connected pieces and after performing a numerical derivative.}
    \label{tab:ensembles_degenerate}
\end{table}
\begin{table}
    \setlength{\tabcolsep}{3pt}
    \rev{
    \begin{tabular}{|c|c|c|c|c|c|c|c|c|c|c|c|c|c|c|}
	\hline
	Ensemble & $\beta$ & $m_0^1$ & $m_0^2$ & $L$ & $T$ & $n_\text{conf}$ & $n_\text{src}$ & $I_{\eta'}$ & $I_{\pi^0}$ & $I_{\pi^\pm}$ & $I_\sigma$ & $I_{\sigma^{\rm conn.}}$ & $I_\rho$ &  $\langle P \rangle$\\
	\hline\hline
	Sp4B1L2M4ND1&6.9&-0.9&-0.89&14&24&300&64&(7, 10)&(7, 10)&(8, 14)&(4, 8)&(8, 14)&(8, 14)&0.55583(5)\\
	Sp4B1L2M4ND2&6.9&-0.9&-0.88&14&24&191&128&(7, 10)&(7, 10)&(8, 14)&(4, 8)&(8, 14)&(8, 14)&0.55474(5)\\
	Sp4B1L2M4ND3&6.9&-0.9&-0.87&14&24&400&128&(7, 10)&(7, 10)&(8, 14)&(5, 8)&(8, 14)&(8, 14)&0.55361(4)\\
	Sp4B1L2M4ND4&6.9&-0.9&-0.85&14&24&300&64&-&-&(8, 14)&(5, 8)&(8, 14)&(8, 14)&0.55163(4)\\
	Sp4B1L2M4ND5&6.9&-0.9&-0.8&14&24&400&128&(7, 10)&(7, 10)&(8, 14)&-&(8, 14)&(8, 14)&0.54735(4)\\
	Sp4B1L2M4ND6&6.9&-0.9&-0.75&12&24&264&64&(7, 10)&(7, 10)&(8, 12)&-&(8, 12)&(8, 12)&0.54395(6)\\
	Sp4B1L2M4ND7&6.9&-0.9&-0.7&12&24&249&128&(7, 10)&(7, 10)&(8, 12)&-&(8, 12)&(8, 12)&0.54104(6)\\
	\hline\hline
\end{tabular}
    }
    \caption{List of all ensembles with non-degenerate fermion masses used in this work. We report the number of configurations $n_{\rm conf}$, the number of the stochastic sources used in the approximation of the all-to-all quark propagator $n_{\rm src}$, the intervals for fitting the resulting meson correlators $I_{\rm meson}$ and the average value of the plaquette $\langle P \rangle$. In some cases we were unable to identify a clear plateau in the effective masses and could not determine the singlet masses. In these cases we do not report a fit interval. For the singlet mesons the interval quoted here was used to fit the correlators after subtracting the excited state contributions in the connected pieces and after performing a numerical derivative.}
    \label{tab:ensembles_non-degenerate}
\end{table}

\subsection{Interpolating operators and two-point functions}\label{ssec:op}
We use fermion bilinear operators for sourcing both singlet and non-singlet mesons. From here onwards, with some abuse of notation, we denote as $\eta'$ and $\sigma$, respectively, the pseudoscalar and scalar, flavor-singlet states. While the mesonic sectors are enlarged in $\Sp{2N}$ with fundamental fermions, every non-singlet or singlet state can still be probed by fermion-antifermion operators, even in the case of non-degenerate fermions. Furthermore, since fermions are moderately heavy, we find such operators are sufficient to study the ground states, and for now do not consider others (such as $\pi\pi$ operators, glueballs, or even tetraquarks). We use the operators
\begin{align}
    \label{eq:operators}
    O_1^{(\Gamma)}(n) &= \bar u(n) \Gamma d(n) \nonumber\,, \\
    O^{(\Gamma)}_\pm(n) &= \left( \bar u(n) \Gamma u(n) \pm \bar d(n) \Gamma d(n) \right) / \sqrt{2}\,,
\end{align}
where $n=(\vec n, t)$ denote lattice sites. For pseudoscalar mesons $\Gamma = \gamma_5$ and we omit the superscript when its value is clear from the context. The pseudoscalar operators $O_-$ and $O_1$ source the pion $5$-plet, and the operator $O_+$ sources the pseudoscalar singlet, $\eta'$.  The same pattern persists for the scalar mesons where $\Gamma = 1$, and we use the notation: 
\begin{align}
    \label{eq:singlet_bilinears}
    O_{\eta'}(n)  &\equiv \left( \bar u(n) \gamma_5 u(n) + \bar d(n) \gamma_5 d(n) \right) / \sqrt{2}\, , \nonumber \\
    O_{\sigma}(n) &\equiv \left( \bar u(n) u(n) + \bar d(n) d(n) \right) / \sqrt{2}\, .
\end{align}
For vector mesons $\Gamma = \gamma_i$ and all operators $O_1$ and $O_\pm$ source states in the vector $10$-plet \cite{Bennett:2019cxd}.   
In the non-degenerate case, the flavored multiplet is always probed by $O_1$. For the vector mesons both $O_-$ and $O_+$ probe the same unflavored multiplet. In the case of the pseudoscalars and scalars, the $O_-$ and $O_+$ probe distinct singlets.
We note that the ensembles studied in this work have moderately heavy fermions---in all cases the vector meson is lighter than twice the pNGB mass. 
After performing the required Wick contractions, we arrive at the following two-point correlation functions. 
\begin{align}
    \label{eq:contraction_diagrams}
    \langle O_{1}(n) \bar O_{1}(m) \rangle  &= - \vcenter{\hbox{\includegraphics[scale=0.5,page=1]{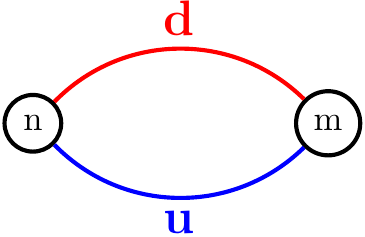}}}  \,,\\
    2\langle O_{\pm}(n) \bar O_{\pm}(m) \rangle   &= - 
    \vcenter{\hbox{\includegraphics[scale=0.5,page=2]{contractions.pdf}}} - \vcenter{\hbox{\includegraphics[scale=0.5,page=3]{contractions.pdf}}} \pm 2~ \vcenter{\hbox{\includegraphics[scale=0.5,page=5]{contractions.pdf}}} +
    \vcenter{\hbox{\includegraphics[scale=0.5,page=6]{contractions.pdf}}} + \vcenter{\hbox{\includegraphics[scale=0.5,page=7]{contractions.pdf}}} \nonumber \,. 
\end{align}
It can be seen that the singlet mesons only differ from the non-singlets by the additional disconnected diagrams. In the degenerate limit they cancel exactly for the $O_-$ operators. In order to determine the mesonic spectrum we need to determine both the connected and disconnected pieces and then fit the zero momentum correlator,
\begin{align}
    C(t) \equiv \sum_{\vec n} \langle O(\vec n, t) \bar O(\vec 0, 0) \rangle\,,
\end{align}
on a Euclidean time interval $(t_{\rm{min}},t_{\rm{max}} )$, where the ground state dominates, and its energy---and thus the mass---can be extracted. The different components of $C(t)$ drop off exponentially with their energy $\propto \exp \left( - E_n t \right)$, and thus at sufficiently large $t$ only the ground state remains, as all other states are exponentially suppressed. However, we note that an additional constant term can arise, which is the case for both the $\eta'$ and the $\sigma$ meson. In the former case this can arise due to an insufficient topological sampling of the path integral, and this constant vanishes in the continuum limit \cite{Aoki:2007ka, Bali:2014pva}. For the scalar singlet, $\sigma$, this constant arises due to the vacuum contributions, e.g.~the fermion condensate, and persists in the continuum limit for vanishing momenta. At large times the correlator $C(t)$ is then given by
\begin{align}
    \label{eq:correlator_constant}
    \lim_{t \to \infty} C(t) = a \left( e^{-m t} + e^{-m (T-t)} \right) + \langle 0 | O | 0 \rangle^2\,,
\end{align}
where the second exponential term is due to the lattice periodicity. While in the case of the $\eta'$ this constant is small compared to the signal and only affects the correlator at large $t$, this is not the case for the $\sigma$ meson. In the scalar case, this constant is several orders of magnitudes larger than the signal and its removal is a significant challenge.

We choose to perform a numerical derivative as proposed in Ref.~\cite{Umeda:2007hy}. The resulting correlator is then anti-symmetric with respect to the midpoint $T/2$, 
\begin{align}
    \label{eq:correlator_derivative}
    \tilde C(t) \equiv \frac{1}{2}\left( C(t-1) - C(t+1) \right) \xrightarrow{t\rightarrow \infty}  a \sinh{(m)} \left( e^{-m t} - e^{-m (T-t)}) \right). 
\end{align}
In order to determine the Euclidean time interval for fitting we use an effective mass $m_{\rm eff}(t) $ defined by
\begin{align}
    \frac{\tilde C(t-1)}{\tilde C(t)} = \frac{e^{ -m_{\rm eff}(t) \cdot (T - t + 1) } \pm e^{ -m_{\rm eff}(t) \cdot ( t - 1) }}{e^{ -m_{\rm eff}(t) \cdot (T - t) } \pm e^{ -m_{\rm eff}(t) \cdot t }}\,,
\end{align}
where the $+$ is used for periodic correlators and the $-$ sign in case of anti-periodic correlators with respect to the lattice midpoint $T/2$. We determine $(t_{\rm min}, t_{\rm max})$ by visually inspecting the effective mass and identifying a plateau at large times $t$. We restrict ourselves to ensembles where the plateau persists over four or more time slices. We then perform a fit of a single exponential term to the correlator $\tilde C(t)$ for the mesons. In Appendix~\ref{sec:direct_subtraction} we compare this method to computing the additional constant $\langle 0 | O | 0 \rangle^2$ directly, without the use of a numerical derivative.

\rev{For the pseudoscalar sector in the non-degenerate $N_f=1+1$ theory both the $\pi^0$ and $\eta'$ are pseudoscalar singlets, and the $\eta'$ is not a groundstate. Thus, we need to perform a variational analysis by computing the correlation matrix of the operators $O_{\pi^0}$ and  $O_{\eta'}$ and solve the resulting generalized eigenvalue problem (GEVP). In the minimal operator basis of \eqref{eq:operators} and \eqref{eq:singlet_bilinears} the cross-correlator are diagrammatically given by
\begin{align}
    2\langle O_{+}(n) \bar O_{-}(m) \rangle   &= - \left( 
    \vcenter{\hbox{\includegraphics[scale=0.5,page=2]{contractions.pdf}}} - \vcenter{\hbox{\includegraphics[scale=0.5,page=3]{contractions.pdf}}} \right)
    +
    \left(
    \vcenter{\hbox{\includegraphics[scale=0.5,page=6]{contractions.pdf}}} - \vcenter{\hbox{\includegraphics[scale=0.5,page=7]{contractions.pdf}}} \right). 
\end{align}
In the mass-degenerate limit the cross-correlator vanishes and the $\eta'$ becomes the ground state of the pseudoscalar singlet sector, whereas the $\pi^0$ becomes part of the pNGB multiplet.
Note the presence of connected diagrams in the cross-correlator. This implies that even in the limit of large fermion masses -- which suppresses the disconnected pieces -- the cross-correlator remains large for large mass differences, i.e. a system with heavy-light properties. Thus, sizeable mixing effects are expected to occur.}

\rev{For a heavy-light system, a more diagonal basis is obtained by using the operators $O^{PS}_{A} = \bar u \gamma u$ and $O^{PS}_{B} = \bar d \gamma d$. The corresponding cross-correlator vanishes as the heavier fermion mass approaches infinity and is given by
\begin{align}
    \langle O^{PS}_{A}(n)  \bar{O}^{PS}_{B}(m) \rangle  &= 
    \vcenter{\hbox{\includegraphics[scale=0.55,page=5]{contractions.pdf}}}.
\end{align}
}

\subsection{Variance reduction techniques}\label{ssec:signalboost}

In order to obtain the full singlet two-point functions we need to measure both the connected and disconnected pieces in Eq.~\eqref{eq:contraction_diagrams}. The disconnected diagrams in particular are very noisy, and the signal is already lost at small to intermediate $t$ where contaminations from excited states are non-negligible. A direct determination of the ground state mass at large $t$ is thus not possible. We can circumvent this problem by removing the contributions of excited states in the singlet correlators manually. This is straightforward for the connected pieces. There, the signal for the connected pseudoscalar and vector mesons persists for all time slices $t$ and in the case of the connected piece of the scalar meson we still have a signal up to large $t$. We fit the connected piece at large times (see Tabs.~\ref{tab:ensembles_degenerate} and \ref{tab:ensembles_non-degenerate} for our choice of fit intervals) to a single exponential 
\begin{align}
    C_{{\rm conn}}^{\rm 1exp}(t) &= A_0 \left( e^{-m_{\rm conn}t} + e^{-m_{\rm conn}(T-t)}  \right),
\end{align}
and replace the full connected piece by the ground state correlator \cite{Neff:2001zr}, where $A_0$ and $m_{\rm conn}$ are the fit parameters, such that 
\begin{align}
 	C_{\eta'}^\text{1exp}(t)  &= C_{\pi,{\rm conn}}^\text{1exp}(t) + C_{\eta',{\rm disc.}}(t)\,, \\
   	C_{\sigma}^\text{1exp}(t) &= C_{\sigma,{\rm conn}}^\text{1exp}(t) + C_{\sigma,{\rm disc.}}(t)\,.
\end{align}
We find that the excited state contributions in the connected pieces are the dominant ones, and removing them shows a much earlier onset of a plateau in the effective masses. 
\rev{The underlying assumption for these correlators is that the excited state contributions of full and connected pieces are indistinguishable within data quality as was noted in \cite{Jansen:2008wv}. This is not guaranteed {\it a priori}. However, we find that the excited state contributions in the connected pieces are indeed the dominant ones.}
In Appendix \ref{sec:smeared_connected} we show that our results obtained by subtracting the connected excited state contributions through a fit at larger times produces the same results as using smeared operators, for the connected pieces with  more overlap with the ground state. 

\rev{Note, that this technique is not applicable to the non-degenerate case, as the $\eta'$ is no longer a ground state and some relevant information is actually encoded in the excited states. Thus, we will not apply this technique there.}

The evaluation of disconnected pieces requires all-to-all propagators. We use $Z_2 \times Z_2$ noisy sources with spin and even-odd dilution \cite{Foley:2005ac}. We typically use  $\mathcal{O} (100)$ distinct noise vectors. The connected pieces are evaluated using stochastic wall sources. Uncertainties are estimated using the jackknife method.

\begin{table}[!t]
    \begin{tabular}{|c|c|c|c|c|c|c|c|c|c|c|}
	\hline
	& $\beta$ & $m_0$ & $L$ & $T$ & $m_\pi L$ & $m_\pi / m_\rho$ & $m_\pi$ & $m_\rho$ & $m_{\eta'}$ & $m_\sigma$\\
	\hline\hline
	SU(2)&2.0&-0.947&20&32&7.47(3)&0.690(7)&0.3735(13)&0.540(5)&-&0.53(4)\\
	SU(2)&2.0&-0.94&14&24&6.40(2)&0.746(6)&0.4576(14)&0.612(4)&0.67(6)&-\\
	SU(2)&2.0&-0.935&16&32&7.91(2)&0.767(5)&0.4946(14)&0.644(4)&0.60(3)&-\\
	SU(2)&2.0&-0.93&14&24&7.491(19)&0.787(4)&0.5350(14)&0.679(3)&0.65(3)&-\\
	SU(2)&2.0&-0.925&14&24&7.999(19)&0.806(4)&0.5713(14)&0.708(3)&0.634(16)&-\\
	SU(2)&2.0&-0.92&12&24&7.323(10)&0.8210(19)&0.6102(8)&0.7432(14)&0.665(9)&-\\
	SU(2)&2.0&-0.9&12&24&8.620(17)&0.862(3)&0.7183(14)&0.832(2)&0.770(16)&-\\
	SU(2)&2.0&-0.88&10&20&8.120(11)&0.885(2)&0.8120(11)&0.9169(19)&0.842(5)&-\\
	\hline\hline
	Sp(4)&7.2&-0.799&32&40&8.087(16)&0.668(4)&0.2527(5)&0.377(2)&-&0.36(5)\\
	Sp(4)&7.2&-0.794&28&36&8.072(11)&0.710(2)&0.2882(4)&0.4055(11)&0.397(16)&-\\
	Sp(4)&7.2&-0.79&24&36&7.505(19)&0.742(6)&0.3127(8)&0.421(3)&0.387(13)&0.56(6)\\
	Sp(4)&7.2&-0.78&24&36&8.882(17)&0.793(4)&0.3700(7)&0.466(2)&0.418(7)&-\\
	Sp(4)&7.2&-0.77&24&36&10.16(2)&0.829(5)&0.4236(10)&0.510(3)&0.456(8)&-\\
	Sp(4)&7.2&-0.76&16&36&7.544(17)&0.850(4)&0.4715(10)&0.554(2)&-&0.64(12)\\
	\hline\hline
	Sp(4)&6.9&-0.924&24&32&8.208(12)&0.663(2)&0.3420(5)&0.5157(17)&-&0.46(3)\\
	Sp(4)&6.9&-0.92&24&32&9.356(12)&0.7036(17)&0.3898(5)&0.5540(12)&-&0.42(2)\\
	Sp(4)&6.9&-0.92&16&32&6.22(2)&0.696(7)&0.3889(14)&0.558(5)&0.49(3)&0.45(6)\\
	Sp(4)&6.9&-0.91&16&32&7.817(19)&0.769(5)&0.4885(12)&0.634(4)&0.560(14)&-\\
	Sp(4)&6.9&-0.91&14&24&6.86(2)&0.766(6)&0.4902(16)&0.639(5)&0.541(9)&0.41(3)\\
	Sp(4)&6.9&-0.9&16&32&9.006(13)&0.815(3)&0.5629(8)&0.690(2)&0.611(9)&-\\
	Sp(4)&6.9&-0.9&14&24&7.897(14)&0.812(3)&0.5641(10)&0.694(2)&0.619(16)&0.57(4)\\
	Sp(4)&6.9&-0.9&12&24&6.796(9)&0.809(2)&0.5663(8)&0.6994(18)&0.610(6)&0.55(2)\\
	Sp(4)&6.9&-0.89&14&24&8.813(19)&0.843(4)&0.6295(13)&0.746(3)&0.69(2)&0.57(9)\\
	Sp(4)&6.9&-0.89&12&24&7.581(15)&0.841(3)&0.6318(12)&0.751(3)&0.661(9)&0.62(7)\\
	Sp(4)&6.9&-0.87&12&24&8.925(10)&0.878(2)&0.7437(9)&0.8468(17)&0.782(13)&0.80(15)\\
	Sp(4)&6.9&-0.87&10&20&7.470(16)&0.871(3)&0.7470(16)&0.857(3)&0.764(9)&-\\
	\hline\hline
\end{tabular}
    \caption{%
The light spectrum  of \SU{2} and \Sp{4} with degenerate fermions: 
the lightest  flavored pseudoscalar, $\pi$, and vector, $\rho$, 
and flavor-singlet pseudoscalar, $\eta^\prime$, and scalar, $\sigma$, mesons, measured in different ensembles.
All dimensionful quantities are given in lattice units 
and the omission of the lattice spacing $a=1$ is understood. 
The missing entries for $m_{\eta'}$ and/or $m_\sigma$
are due to the measurements 
 not fulfilling the fitting criteria discussed in the main text. 
    }%
    \label{tab:masses_degenerate}
\end{table}

\section{Results}\label{sec:results}

Here we report the main results of our numerical investigations on 
the mass spectrum of flavor-singlet pseudoscalar and scalar mesons, obtained using 
the techniques discussed in the previous section. 
We focus on the $\Sp{4}$ theory coupled to two fundamental dynamical fermions, 
but for degenerate fermions 
we supplement it with the $\SU{2}$ theory with the same matter content. 
In the case of the pseudoscalar singlet,
we further compare to the existing literature on lattice results 
for the $\SU{3}$ theory. 

\begin{figure}
    \includegraphics[width=0.49\textwidth]{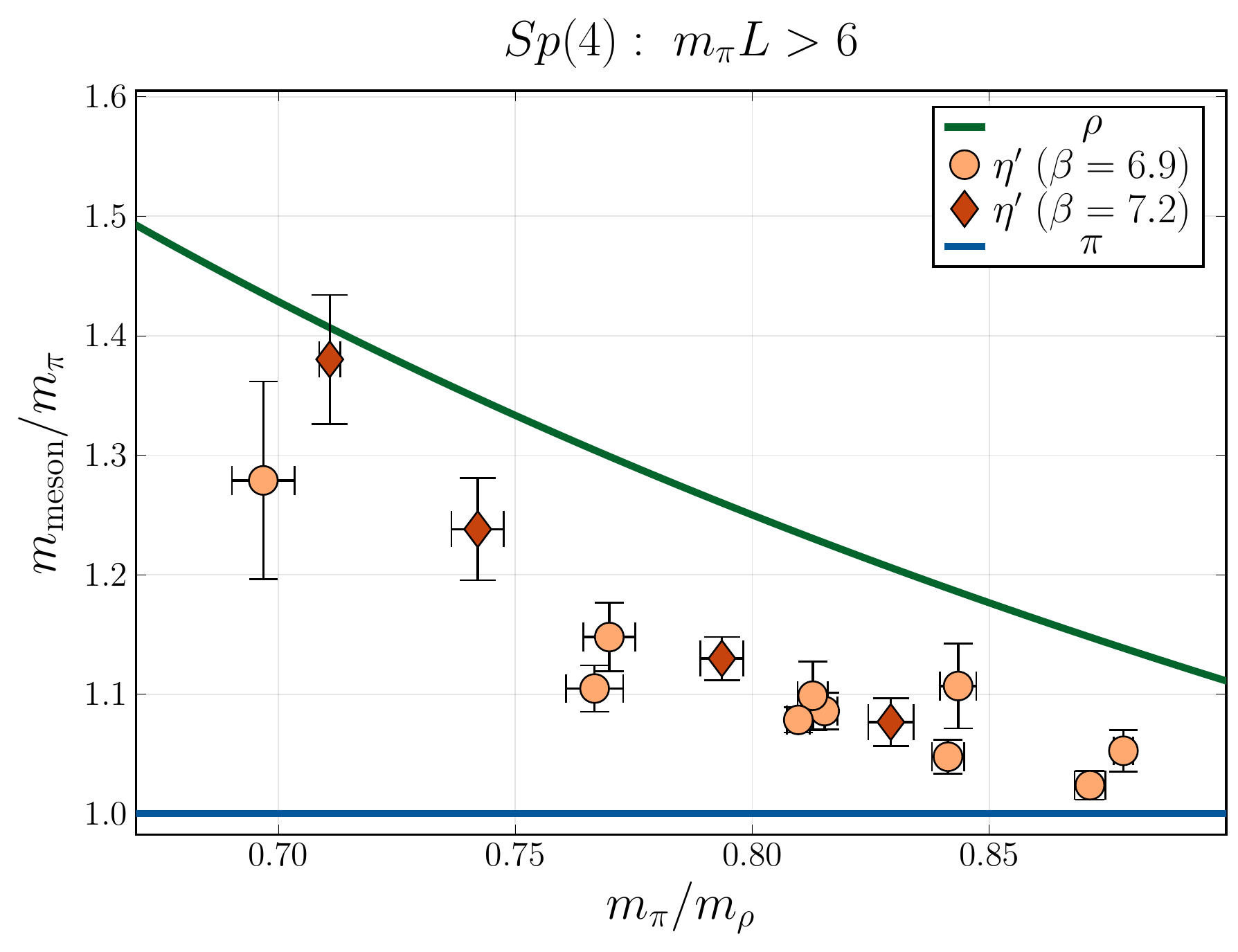}
    \includegraphics[width=0.49\textwidth]{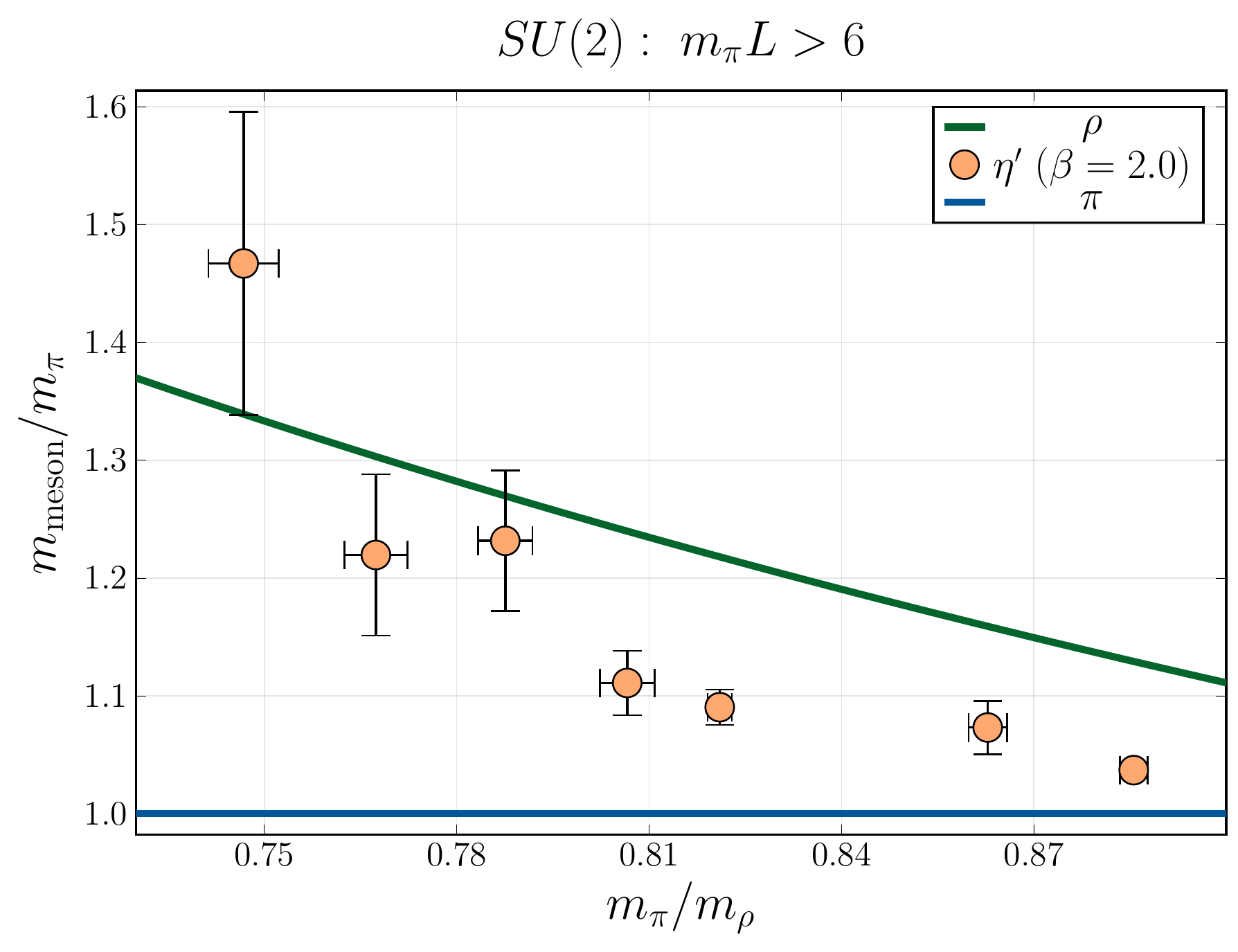}
    \caption{%
    (left panel) Mass ratios $m_{\rm meson}/m_\pi$ for pseudoscalar and vector mesons, including the flavor-singlet pseudoscalar $\eta^\prime$, in the $\Sp{4}$ gauge theory with $N_f=2$ Dirac flavors of fermions in the fundamental representation measured at the two values of the inverse coupling $\beta=6.9$ and $7.2$. (right panel) The same plot but in the $\SU{2}$ gauge theory at $\beta=2.0$. The green solid lines $m_\rho/m_\pi = 1/x$ are displayed for reference. 
    }%
    \label{fig:eta_deg}
\end{figure}

\subsection{Pseudoscalar singlet in \SU{2} and \Sp{4} with \texorpdfstring{$N_f=2$}{Nf=2}}\label{ssec:psdeg}

Our results for the mesons with degenerate fermions are tabulated in Tab.~\ref{tab:masses_degenerate}. 
All the ensembles satisfy the condition $m_\pi L > 6$,
suggested by the observations in Ref.~\cite{Bennett:2019jzz} for $\Sp{4}$, and in Ref.~\cite{Hietanen:2014xca} for $\SU{2}$, that
the size of finite volume corrections to the low-lying spectrum 
for flavored mesons is 
of the order of $1\sim 2\%$ at $m_\pi L \simeq 6$, 
and becomes much smaller for the larger volumes, 
as it is exponentially suppressed with the volume.
This observation is also confirmed by our  measurements 
of $m_\pi$ and $m_\rho$ at different volumes in the $\Sp{4}$ theory 
with $\beta=6.9$, by varying the bare fermion mass, $m_0$. 
Finite volume corrections to $m_{\eta^\prime}$ are 
compatible with the statistical uncertainties 
and expected to be less than $2\%$, which we
estimated from the most precise results available, for $m_0=-0.9$, if $m_\pi L \gtrsim 6$. 
We therefore safely neglect finite volume corrections to $m_{\eta^\prime}$ in the following. 

In Fig.~\ref{fig:eta_deg}, we present our measurements of the ratios between the mass of 
the $\eta^{\prime}$ meson 
and that of the pseudoscalar non-singlet $\pi$, as a function of $m_\pi/m_\rho$. For reference, we indicate the mass of the vector meson $\rho$ by a solid line. 
In the $\Sp{4}$ theory 
we find that the pseudoscalar singlet is consistently heavier than the non-singlet, 
over the range of $0.7 \lesssim m_\pi/m_\rho \lesssim 0.9$, 
but lighter than the vector mesons. 
While in the lightest and finest ensembles the hierarchy between the pseudoscalar singlet and vector mesons 
is not yet clearly resolved,
the emerging trend is that $m_{\eta^\prime}/m_\pi$ slowly increases 
as $m_\pi$ decreases in this  mass regime, and approaches 
$m_\rho/m_{\pi}$ for $m_\pi/m_\rho \lesssim 0.75$. 
We do not observe an appreciable difference 
in the mass ratios obtained with the two different values of $\beta$, within the quoted one-sigma error bars. 
We find a similar trend 
in the $\SU{2}$ theory, as shown in the right panel of Fig.~\ref{fig:eta_deg}. 
Since in this case only one, fairly coarse lattice is considered, 
we cannot comment on the size of finite lattice spacing effects. 

\begin{figure}
    \includegraphics[width=0.49\textwidth]{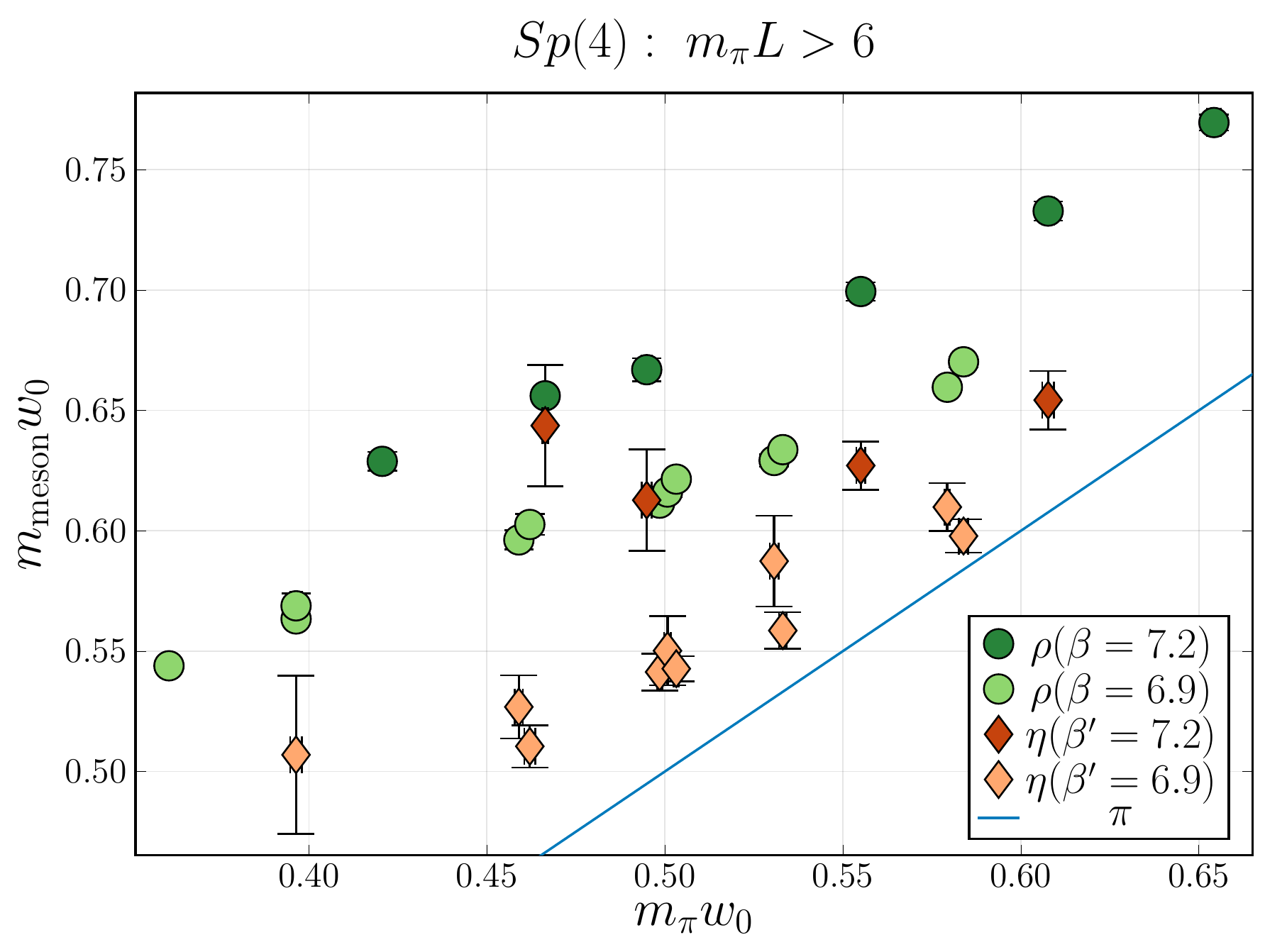}
    \includegraphics[width=0.49\textwidth]{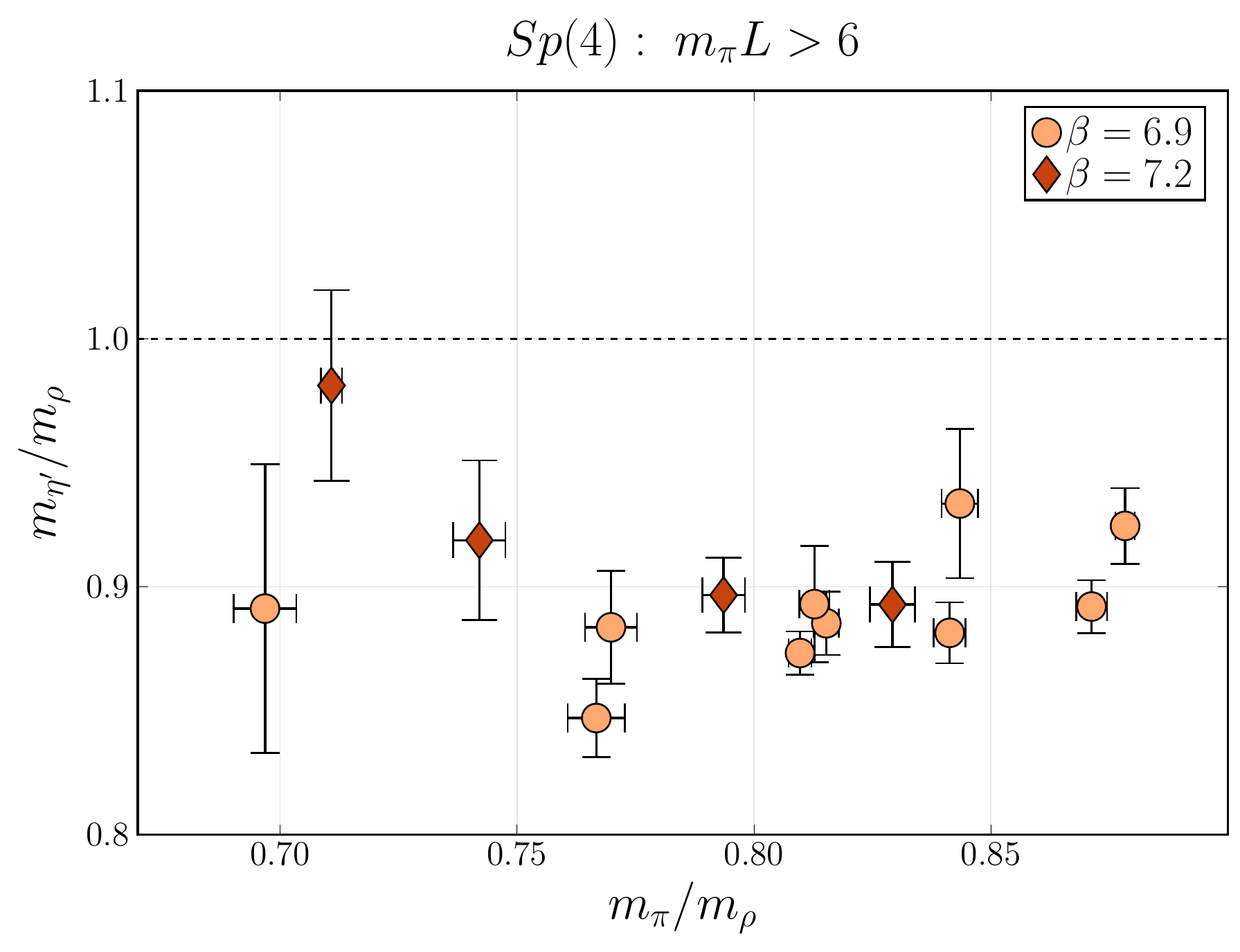}
    \caption{%
    (left panel) Same data as in the left panel of Fig.~\ref{fig:eta_deg}, but now masses are given in units of the gradient flow scale $w_0$. The blue solid line $w_0 m_\pi = x$ is displayed for reference. (right panel) Mass ratio between of the vector mesons $\rho$ and the scalar singlet $\eta'$. 
    }%
    \label{fig:eta_deg_GF}
\end{figure}

The smallness of  lattice artifacts in the ratios of meson masses is somewhat surprising, 
as the lattice spacing for $\beta=7.2$ is approximately $40\%$ smaller than for $\beta=6.9$~\cite{Bennett:2019jzz}. 
To assess this point, we present the meson masses in units of 
 the gradient flow scale $w_0$,
which defines a common scale  in the continuum theory,
and which we use also  
to compute the topological charge $Q$---see Appendix~\ref{sec:direct_subtraction}.
We borrow the definition and measurements of the gradient flow scale $w_0$ from Ref.~\cite{Bennett:2019jzz}, and refer the reader to that publication for details.
The left panel of Fig.~\ref{fig:eta_deg_GF} shows that both
the mass of the pseudoscalar singlet 
and the vector mesons 
receives
significant corrections from the finite lattice spacing. 
By comparing with Fig.~\ref{fig:eta_deg}, we see that such corrections to $m_\pi w_0$ and $m_{\eta^\prime} w_0$ happen to have the same sign and similar sizes, 
which cancel out in the mass ratios. We observe the same pattern for the mass ratio of $m_\rho$ and $m_{\eta^\prime}$, as depicted in the right panel of Fig.~\ref{fig:eta_deg_GF}.

\subsection{Pseudoscalar singlets in Sp(4) with \texorpdfstring{$N_f=1+1$}{Nf=1+1}}\label{ssec:psnondeg}
\begin{table}
    \setlength{\tabcolsep}{3pt}
    \rev{
    \begin{tabular}{|c|c|c|c|c|c|c|c|c|c|c|c|}
	\hline
	$\beta$ & $m_0^{(1)}$ & $m_0^{(2)}$ & $L$ & $T$ & $m_{\pi^0}$ & $m_{\pi^0_c}$ & $m_{\pi^\pm}$ & $m_{\rho^0}$ & $m_{\rho^\pm}$ & $m_{\eta'}$ & $m_{\sigma}$\\
	\hline\hline
	6.9&-0.9&-0.89&14&24&0.60(2)&0.597(2)&0.596(2)&0.723(3)&0.719(4)&0.65(4)&0.45(6)\\
	6.9&-0.9&-0.88&14&24&0.63(3)&0.6261(18)&0.628(2)&0.749(2)&0.745(3)&0.69(4)&0.51(8)\\
	6.9&-0.9&-0.87&14&24&0.664(17)&0.6510(14)&0.6606(17)&0.7731(19)&0.775(2)&0.77(3)&0.69(14)\\
	6.9&-0.9&-0.85&14&24&-&0.6889(19)&0.709(2)&0.813(3)&0.813(4)&-&0.51(12)\\
	6.9&-0.9&-0.8&14&24&0.75(2)&0.7563(16)&0.8277(18)&0.879(2)&0.922(3)&0.93(2)&-\\
	6.9&-0.9&-0.75&12&24&0.79(3)&0.803(4)&0.921(4)&0.927(6)&1.005(5)&1.04(3)&-\\
	6.9&-0.9&-0.7&12&24&0.83(3)&0.833(4)&0.996(3)&0.954(5)&1.073(5)&1.15(3)&-\\
	\hline\hline
\end{tabular}
    \caption{%
    Results for the light spectrum with non-degenerate fermions: the flavored pseudoscalar $\pi^\pm$ and vector $\rho^\pm$, and the flavor-singlet pseudoscalar $\eta^\prime$, $\pi^0$, vector $\rho^0$ and scalar $\sigma$ mesons. 
    All dimensionful quantities are given in lattice units 
    ($a=1$).
    Missing entries for $m_{\eta'}$ or $m_\sigma$ denote measurements that do not fulfill the fitting criteria discussed in the main text. We furthermore report the mass of the flavor-singlet pseudo-Goldstone $m_{\pi^0_c}$ in the connected-only approximation. 
    }}%
    \label{tab:masses_non-degenerate}
\end{table}

\begin{figure}
    \includegraphics[width=0.49\textwidth]{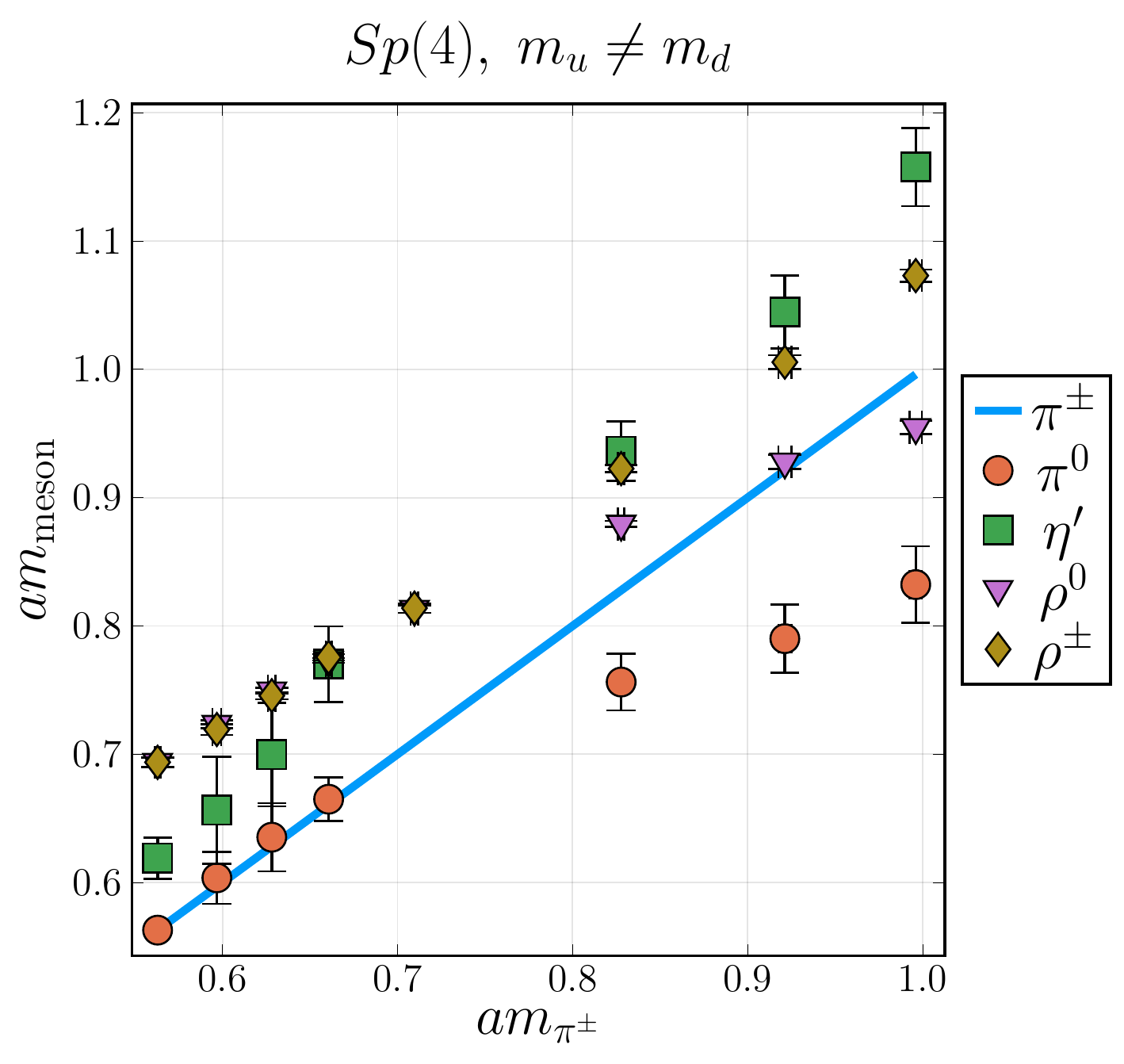}
    \includegraphics[width=0.49\textwidth]{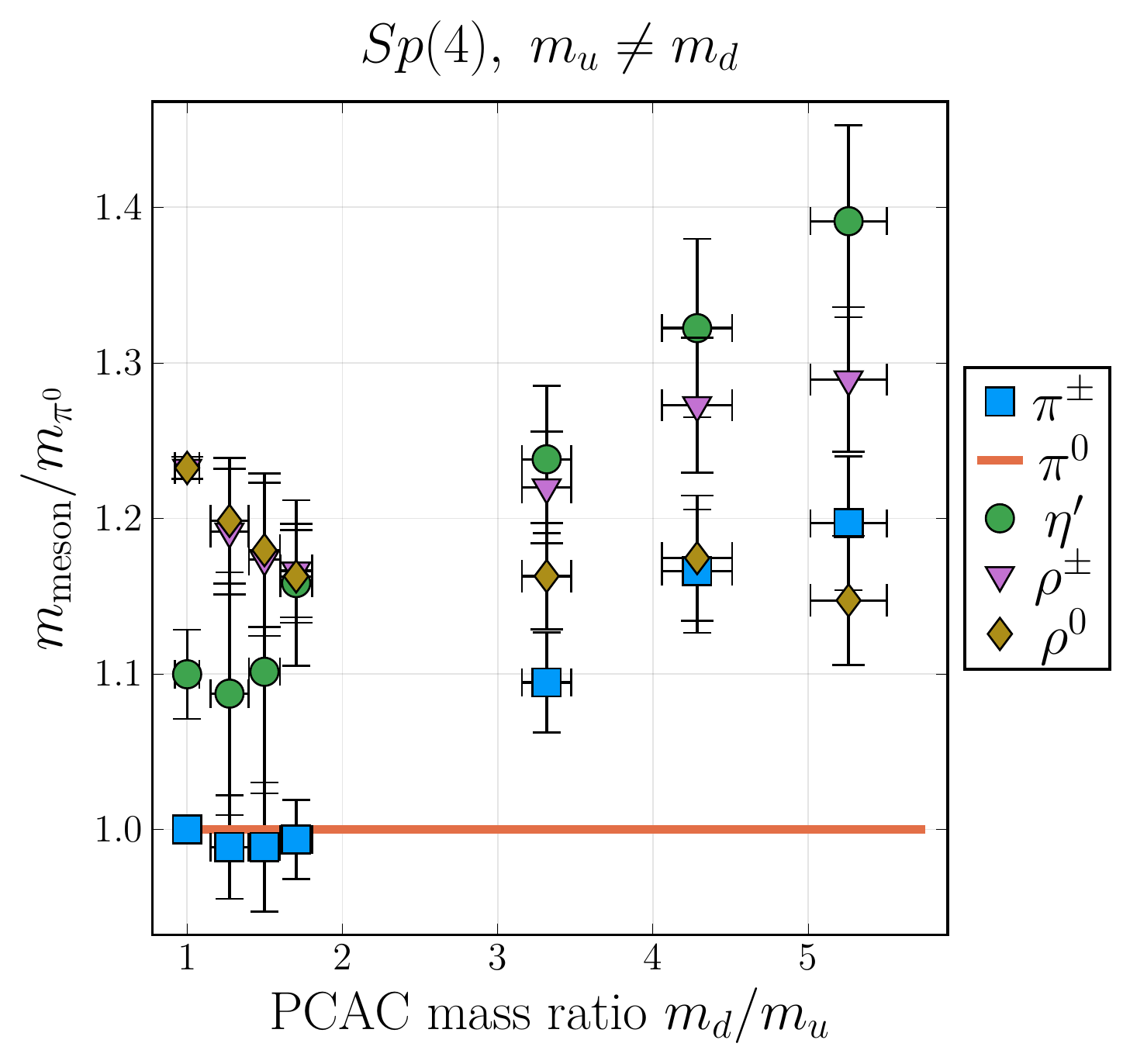}
    \caption{\rev{
    Masses of the lightest non-singlet mesons as well as the pseudoscalar singlet meson in the $\Sp{4}$ theory with non-degenerate Dirac fermions. We fix the lattice coupling and one of the bare fermion masses to $\beta=6.9$ and $m_0^{(1)}=-0.9$, respectively, while varying the other bare fermion mass. In the left panel we display our results as a function of the flavoured pion mass, while in the right panel we show them as a function of ratio of the PCAC fermion masses in units of the $\pi^0$ mass.
    }}
    \label{fig:eta_nondeg}
\end{figure}

For non-degenerate fermions,  the theory contains two flavor-singlet pseudoscalar 
mesons, the $\eta^\prime$ as well as the flavor-diagonal PNGB, $\pi^0$. 
To understand the effects of (explicit) flavor-symmetry breaking 
on the low-lying spectrum, we first choose the ensemble for degenerate fermions 
with $\beta=6.9$ and $m_0=-0.9$ and vary the bare mass of one of the Dirac fermion, $m_0^{(2)}\geq m_0$, for which we effectively increase its mass, while keeping that of the other fixed, $m_0^{(1)} = m_0$. 
We summarize the numerical results in Tab.~\ref{tab:masses_non-degenerate}. 
In the table, we also present the mass of the flavor-singlet PNGB 
obtained by computing only the connected diagrams after dropping the last three terms in Eq.~\ref{eq:contraction_diagrams}, which we denote by $m_{\pi_c^0}$. 
\rev{Within the generally larger uncertainties}, we find no statistically appreciable difference between $m_{\pi^0}$ and $m_{\pi_c^0}$, which supports the connected-only approximation considered in Ref.~\cite{Kulkarni:2022bvh}. 

\rev{In the left panel of Fig.~\ref{fig:eta_nondeg}, we show the meson masses, as a function of the flavoured pNGBs mass $m_{\pi^\pm}$. In the mass-degenerate limit, we recover the mass hierarchy of Sec.\ref{ssec:psdeg}, as expected. As we increase one of the fermion masses, we observe a clear separation between flavored mesons and the unflavoured $\pi^0$ as well as the unflavoured vector mesons $\rho^0$, with the former being heavier than the latter.} 
\rev{At the same time, the pseudoscalar singlet $\eta^\prime$, becomes heavier. This effect leads to an inversion of the mass hierarchy between the vector mesons $\rho^\pm$, $\rho^0$ and the $\eta^\prime$ meson and the $\eta'$ becomes heavier than any other meson considered here. \footnote{\rev{This is qualitatively different from the preliminary results in \cite{Zierler:2022qfq}, where the mixing between the pseudoscalar singlets was neglected and the excited state subtraction of Sec.~\ref{ssec:signalboost} was incorrectly applied to the non-degenerate theory.}}}

A couple of cautionary remarks should be added. First of all, we are in a moderately heavy mass regime, with $m_{\pi^0}/m_{\rho^0}\sim 0.85$. Secondly, some meson masses for heavy ensembles sit close to the lattice cut-off and thus could be affected by significant lattice artifacts. 

\rev{For the heaviest fermion masses, the mass difference between the $\eta'$ and the $\pi^0$ is approximately twice the mass difference between the $\pi^0$ and the $\pi^\pm$'s. This indicates that the mass-differences are driven by the valence fermion masses.} 

\rev{In this regime, the singlet $\pi^0$ and the non-singlet $\rho^0$ approach an approximate one-flavour theory} --- see e.g.\  Ref.~\cite{Francis:2018xjd} for lattice results on the low-lying spectrum in the $\SU{2}$ gauge theory with one Dirac fermion. 

In the right panel of Fig.~\ref{fig:eta_nondeg} we plot the meson masses as a function of the quark mass ratio, defined through the partially conserved axial current (PCAC) relation. We identify the average quark mass in the flavored pion $\pi^\pm$ through the relation
\begin{align}
\label{eq:PCAC_masses}
m^{\rm PCAC}_{\rm avg.} = \lim_{t \to \infty} \frac{1}{2} \frac{\partial_t C_{\gamma_0 \gamma_5, \gamma_5}(t)}{C_{\gamma_5}(t)} = \lim_{t \to \infty} \frac{1}{2} \frac{\partial_t \int d^3\vec x \langle  \left( \bar u (\vec x, t) \gamma_0 \gamma_5 d (\vec x, t) \right)^\dagger \bar u (0)  \gamma_5 d (0)  \rangle}{\int d^3\vec x \langle  \left( \bar u (\vec x, t)\gamma_5 d (\vec x, t) \right)^\dagger \bar u (0)  \gamma_5 d (0)  \rangle}. 
\end{align}
The unrenormalized PCAC quark mass ratio $m_d/m_u$ for non-degenerate fermions is extracted by performing an additional measurement  of the PCAC average mass at degeneracy. For a detailed discussion of the PCAC relation and different ways to calculate it, we refer to Ref.~\cite{Gattringer:2005ij}.

\subsection{Scalar singlet in Sp(4) with \texorpdfstring{$N_f=2$}{Nf=2}}\label{ssec:sdeg}
\begin{figure}
    \includegraphics[width=0.60\textwidth]{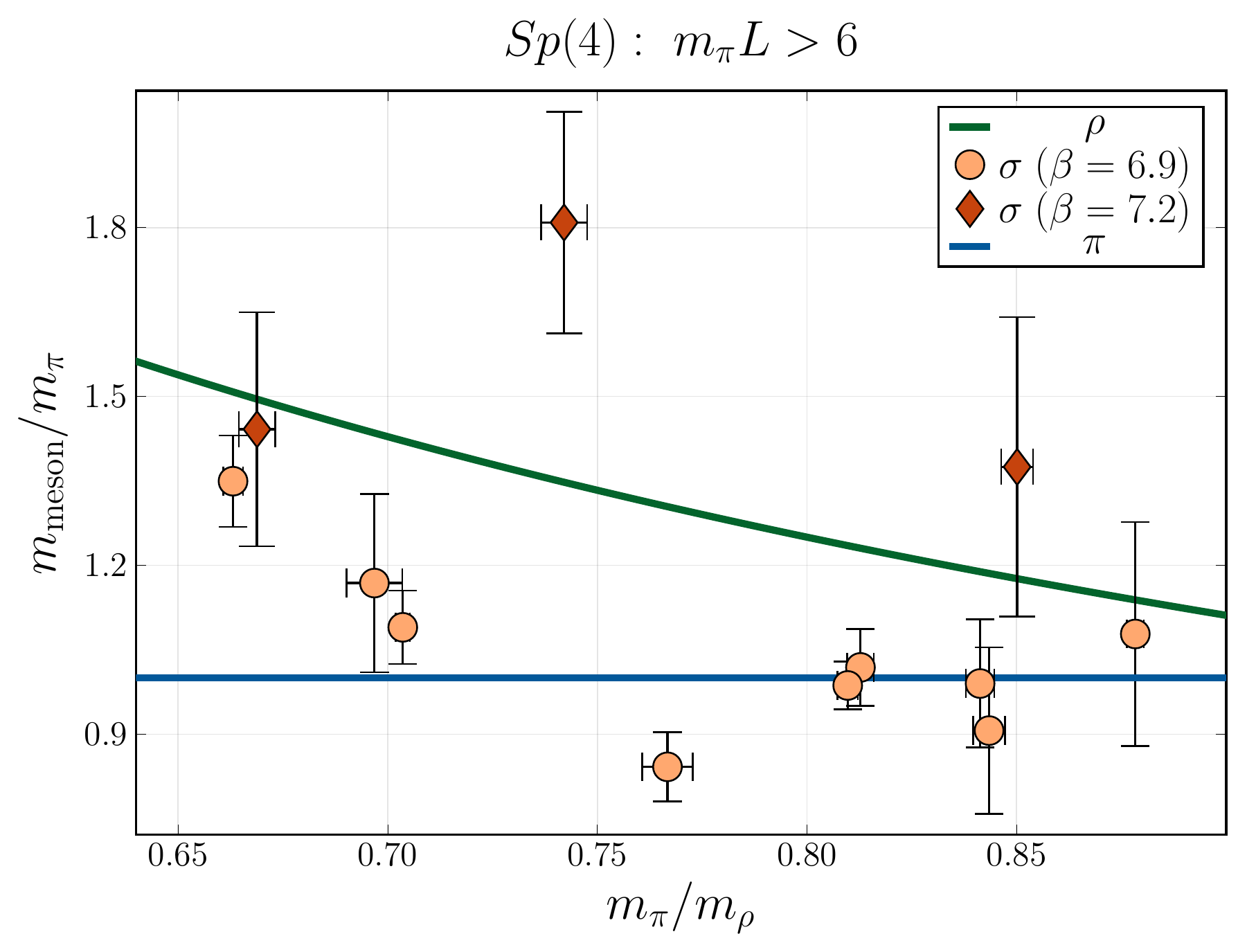}
    \caption{Mass of the $\sigma$ meson with degenerate fermions in the $\Sp{4}$ theory. We find signs of finite spacing effects even when considering  ratios of hadron masses. On the coarser lattice the scalar singlet appears to be quite light, in some cases even lighter than the $\pi$. For the finer lattice this changes drastically and the scalar singlet $\sigma$ is usually heavier than the vector meson $\rho$, though still below the two-$\pi$ threshold. The green solid line $m_\rho/m_\pi = 1/x$ is displayed for reference. }
    \label{fig:sigma_deg}
\end{figure}

In the case of the scalar singlet, $\sigma$, the signal is consistently worse than for the other states discussed so far. Furthermore, we see signs of finite spacing effects, shown in Fig.~\ref{fig:sigma_deg}. For $\Sp{4}$ on the coarse $\beta=6.9$ lattices, we observe a light $\sigma$ state, of mass comparable to the mass of the $\pi$. This pattern persists over the entire mass range considered. On finer lattices, for $\beta=7.2$, the mass of the $\sigma$ increases and is heavier than the PNGBs, and comparable to the vector meson, though with much larger statistical errors, and  still below the $\pi\pi$ threshold.

These results suggest the existence of larger finite-spacing effects that affect the mass of the scalar singlet. Yet, some caution should be used, because, due to the  large noise in our signal, the mass is extracted from much shorter times on the finer lattices, and may therefore also be more severely  affected by excited-state contamination, and possibly other systematics.
Nevertheless, even for the finer lattice, the $\sigma$ state is lighter than its non-singlet counterpart, suggesting that further studies are still needed. The scalar singlet might be a stable light meson at moderately heavy fermion masses, and thus phenomenologically relevant.

\subsection{Comparison to SU(3) with \texorpdfstring{$N_f=2$}{Nf=2}}\label{ssec:compsu3}
\label{sec:su3_eta}

\begin{figure}
    \includegraphics[width=0.99\textwidth]{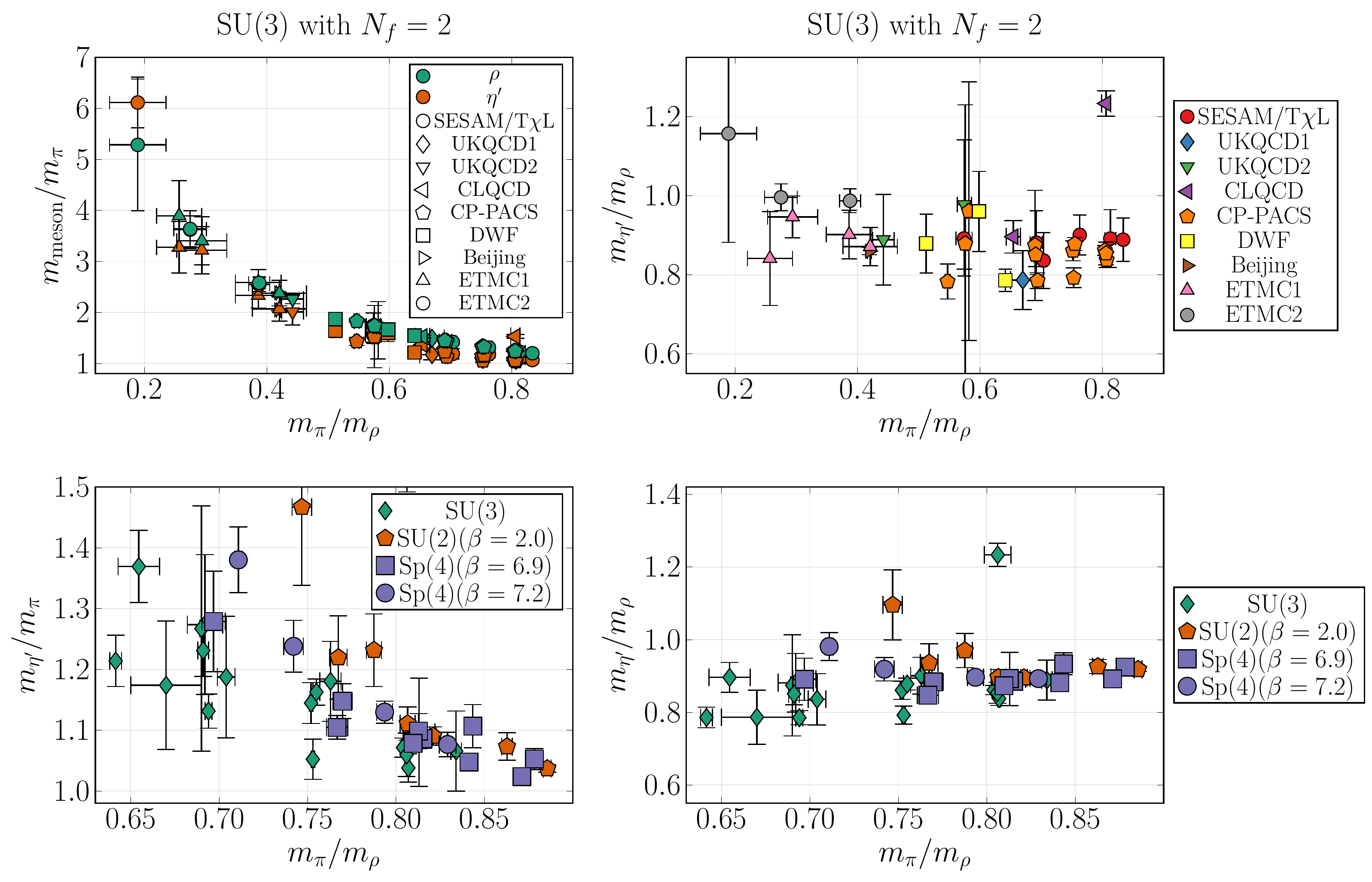}
    \caption{Comparison to the available lattice data in $\SU{3}$ with two fundamental fermions. The upper panels depict all available lattice results in $\SU{3}$. In the upper left panel the green markers denote the vector meson $\rho$ and the orange ones the pseudoscalar singlet $\eta'$. The different marker shapes denote the different collaborations. We see that the $\eta'$ is lighter than the vector mesons almost everywhere. In the upper right panel we directly plot the ratio $m_{\eta'} / m_\rho$. In the lower panels we compare the $\SU{3}$ results to our $\Sp{4}$ and $\SU{2}$ data. In the lower left panel, we show the ratio $m_{\eta'}/m_\pi$ as a function of $m_\pi / m_\rho$ for values of $m_\pi / m_\rho \approx 0.7$ and larger. In the lower right plot we compare the different results of the ratio $m_{\eta'}/m_\rho$.}
    \label{fig:SU3_comparison}
\end{figure}

In Fig.~\ref{fig:SU3_comparison} we show a compilation of the available data published on the pseudoscalar singlet for the $\SU{3}$ theory with $N_f=2$ (upper panels) as well a comparison of our results for $\Sp{4}$ and $\SU{2}$ to the available data for $\SU{3}$ (lower panels). In some cases, the measurement has been performed using different methods in the analysis or different operators have been used to study the same mesons (e.g.\ the mass of the $\eta'$ has been obtained from pure gluonic operators as well as the usual fermionic operators, or in the case of twisted mass fermions the non-singlet mesons include isospin breaking effects) and sets of results are available. In such cases, we have chosen the results that are closest to the standard determination of directly fitting the correlator of a pure fermionic operator. When this was not possible we quote the largest and smallest values of $ m_{i} \pm \Delta m_{i}$ of all measurements $i$ and symmetrize the uncertainties.  
The data depicted in Fig.~\ref{fig:SU3_comparison} has been taken from the UKQCD collaborations (denoted by UKQCD1~\cite{UKQCD:1998zbe,McNeile:2000hf} and UKQCD2~\cite{Allton:2004qq}); the SESAM/T$\chi$L collaboration~\cite{TXL:2000mhy,Bali:2000vr}; the CP-PACS collaboration~\cite{CP-PACS:2002exu}; the RBC collaboration using domain-wall fermions~\cite{Hashimoto:2008xg}; the CLQCD collaboration using Wilson clover fermions on anisotropic lattices~\cite{Sun:2017ipk}; the ETMC collaboration (denoted by ETMC1~\cite{Urbach:2007rt,Jansen:2008wv} and ETMC2~\cite{Dimopoulos:2018xkm,Fischer:2020yvw}); and from the analysis of $\eta'$-glueball mixing (denoted by Beijing~\cite{Jiang:2022ffl}). 

In all but the very lightest ensemble (and one obvious outlier at heavy fermion mass) the vector meson, $\rho$, is found to be heavier than the pseudoscalar singlet, $\eta'$. The authors of Ref.~\cite{Fischer:2020yvw} point out that in the lightest ensemble the $\rho$ particle might be unusually light due to the small number of energy levels below the inelastic threshold in the determination of the $\pi\pi$ phase shift. It is lighter than their extrapolation to the physical point at which $m_\rho = 786(20)$ and even lighter than their extrapolation to the chiral limit. The mass dependence of the $\eta'$ meson was found to be flat and an extrapolation in Ref.~\cite{Dimopoulos:2018xkm} to the physical point gave $m_{\eta'} = 772(18) \rm{MeV}$. This is in contrast to SM QCD where the $\eta'$ is significantly heavier---the current PDG lists $m_{\eta'}^{\rm{PDG}} = 957.78(6) \rm{MeV}$~\cite{ParticleDataGroup:2022pth}, which is in agreement with recent $\SU{3}$, $N_f=2+1$ lattice results of $m_{\eta'} = 929.9  \left(\substack{47.5\\21.0}\right)$~\cite{Bali:2021qem}. This suggests it is the contribution of the s-quark that leads to the heavier mass. This can be understood in a quark model of the pseudoscalar singlet mesons based on approximate $\SU{3}_F$ flavor symmetry \cite{Feldmann:1998sh,Feldmann:1998vh,Feldmann:1999uf} which was applied to early lattice results in \cite{McNeile:2000hf}.

The bottom line of this brief survey is that in the regime of moderately large fermion masses the pattern of ground state masses observed so far in $SU(3)$ is quite similar, both qualitatively and quantitatively, to our findings in the \Sp{4} case as can be seen in the lower panels of Fig.~\ref{fig:SU3_comparison}. The gauge group and modified chiral structure do not seem to have a very strong impact on mass of the $\eta'$.

\subsection{Possible phenomenological implications}

Our results provide evidence that the singlet sector, computed for moderately large fermion masses in the \Sp{4} theory, is not dissimilar from what is observed in the $\SU{2}$ and  $\SU{3}$ theories coupled to two fundamental fermions.
In particular, both pseudoscalar and scalar singlets are light enough to be stable against decay into Goldstone bosons, over a fermion-mass range within which also the flavored vector mesons would not decay. We now present a few examples of potential implications for phenomenological models for which these theories can be invoked to yield a short-distance completion.

Firstly, because the flavor singlets are not much heavier than the flavored mesons, if a model of this type were used as  part of a hidden valley scenario, or a new dark sector, these states would then only decay via a mediator mechanism into standard model particle, but not strongly, and would be long-lived. Their lifetime and branching fractions would be determined by the detailed structure of the coupling to the standard-model fields.
They  are unlikely to be long-lived enough to play a significant role in a model explaining current  dark matter density, yet they can easily appear as long-lived particles in experiments~\cite{Albouy:2022cin,Blondel:2022qqo,Cottin:2022nwp,ATLAS:2023oti}, and hence the existence of a new dark sector containing this theory is experimentally testable. 

Other possible observable effects in this context could arise because the singlets can enhance interaction cross-sections, as virtual particles, affecting processes even below production threshold. They can therefore play a relevant role for dark matter self-interactions~\cite{Kaplinghat:2015aga}. Their effect could even affect form factors relevant to direct detection experiments~\cite{Laha:2013gva}. Depending on details, they could also offer a possibility to create indirect detection signatures in cases of high dark matter densities. Finally, both singlets can serve, together or individually, as a Higgs portal, removing the need for an independent messenger.

In the alternative context of composite Higgs scenarios, in which the PNGBs provide the longitudinal components of  the $W$-bosons and $Z$ boson, as well giving rise to the experimentally observed
Higgs boson, the pseudoscalar singlet can become a surprisingly strong limiting factor~\cite{Andersen:2011yj}. As its signature is possibly similar to that of the pseudoscalar Higgs in the minimal supersymmetric standard model, or in classes of  two-Higgs doublet models, strong exclusion limits  already exist, both for a pseudoscalar Higgs heavier and lighter than the standard model Higgs. To avoid  these bounds requires to open up substantially the mass gap between the scalar  and pseudoscalar singlets, but in our measurements we always observe the opposite hierarchy.

\section{Summary}\label{sec:summary}

We have presented the results of the first dedicated lattice study of flavor singlet meson states in the \Sp{4} theory coupled to two (Wilson-Dirac) fundamental, dynamical fermions. We have computed the masses of the lightest pseudoscalar and  scalar singlets in a portion of parameter space in which the fundamental  fermion are moderately heavy. We have considered both the case of degenerate and of non-degenerate masses for the fermions. The continuum limit of this theory, in the range of parameters explored, is of interest because it provides the ultraviolet completion of several proposals for  new physics extensions of the standard model, in the contexts of composite Higgs models and strongly interacting dark matter. In order to perform this study, we implemented in our analysis state-of-the-art techniques to account for the contribution of disconnected diagrams to correlation functions involving flavor singlets.

We observe that the qualitative (and to large extent even the quantitative) features of the mass spectrum we find in this \Sp{4} theory are similar to those of \SU{2} and \SU{3} theories with the same field content, in comparable  ranges of parameter space. More specifically,  the mass range of the singlet states, in particular  of the lightest pseudoscalar, is comparable
 to the masses of the lightest flavored mesons, at least for our choices of fermion masses. 
 This remains true also in the mass-non-degenerate case.
 
Our findings suggest that the singlet sector cannot be neglected in phenomenological studies of models that have their dynamical, short-distance origin in this theory.  However, notwithstanding the technical implementation of several techniques to enhance the signal-to-noise ratio in our measurements, and the comparatively large statistics provided by our numerical ensembles, we have also found that the observables are affected by large lattice artifacts, especially in the case of the scalar singlet. While we have noticed that taking certain ratios of masses reduces drastically the size of such effects, if phenomenological considerations require precision measurements for the mass spectrum, then this would provide strong incentive to further improve this study, in particular in order to better understand the approach to the continuum limit.

On more general and abstract theoretical ground, the similarity of our main results with the \SU{N} cases strongly suggests that the altered chiral structure and gauge group has limited impact on the underlying dynamics. On the one hand, this might be expected in a gauge theory with small number of moderately heavy fermions. On the other hand, though, by extending this kind of analysis to different $N$ and/or further gauge groups we envision to be able to gain quantitative understanding the relevance of gauge dynamics for hadron dynamics beyond group-theoretical, and thus non-dynamical, aspects.

\begin{acknowledgments}
We are grateful to S.~Kulkarni for helpful discussions and to the authors of \cite{Lee:2022elf,Hsiao:2022kxf} for providing access to the smearing code in HiRep prior to publication. 

FZ is supported by the Austrian Science Fund research teams grant STRONG-DM (FG1) and acknowledges travel support for part of this work by the city of Graz. The work of EB is supported by the UKRI Science and Technology Facilities Council (STFC) Research Software Engineering Fellowship EP/V052489/1, and by the ExaTEPP project EP/X017168/1.
The work of JWL was supported in part by the National Research Foundation of Korea (NRF) grant funded 
by the Korea government(MSIT) (NRF-2018R1C1B3001379) and by IBS under the project code, IBS-R018-D1. 
The work of BL and MP has been supported in part by the STFC 
Consolidated Grants No.~ST/P00055X/1 and No. ST/T000813/1.
BL and MP received funding from
the European Research Council (ERC) under the European
Union’s Horizon 2020 research and innovation program
under Grant Agreement No.~813942. 
The work of BL is further supported in part 
by the Royal Society Wolfson Research Merit Award 
WM170010 and by the Leverhulme Trust Research Fellowship No. RF-2020-4619.
The work of HH is supported by the Taiwanese MoST grant 109-2112-M-009-006-MY3.

The computations have been partially performed on the HPC cluster of the University of Graz and on the Vienna Scientific Cluster (VSC4), and partially on the DiRAC Data Intensive service at Leicester. The DiRAC Data Intensive service at Leicester is operated by the University of Leicester IT Services, and forms part of the STFC DiRAC HPC Facility (www.dirac.ac.uk). The DiRAC Data Intensive service equipment at Leicester was funded by BEIS capital funding via STFC capital grants ST/K000373/1 and ST/R002363/1 and STFC DiRAC Operations grant ST/R001014/1. DiRAC is part of the National e-Infrastructure.

\vspace{1.0cm}

{\bf Open Access Statement}---For the purpose of open access, the authors have applied a Creative Commons 
Attribution (CC BY) licence  to any Author Accepted Manuscript version arising.

\vspace{1.0cm}

{\bf Research Data Access Statement}---The data generated for this manuscript can be downloaded from Ref.~\cite{data_release}, and the software used to analyze and present it is similarly available from Ref.~\cite{analysis_release}. 

\end{acknowledgments}

\appendix
\section{Constant contributions to the correlators}
\label{sec:direct_subtraction}

\begin{figure}
    \includegraphics[width=.49\textwidth]{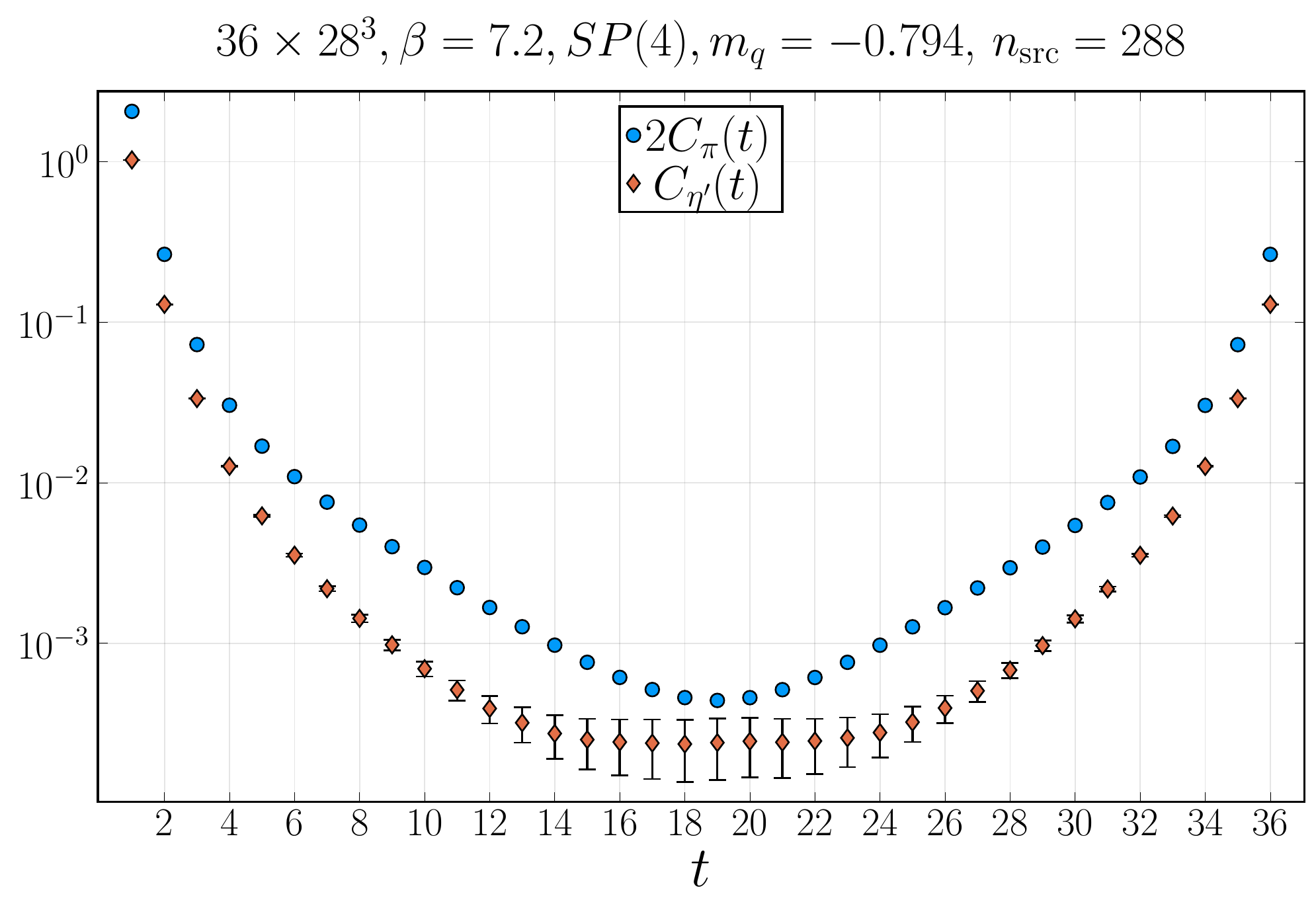}
    \includegraphics[width=.49\textwidth]{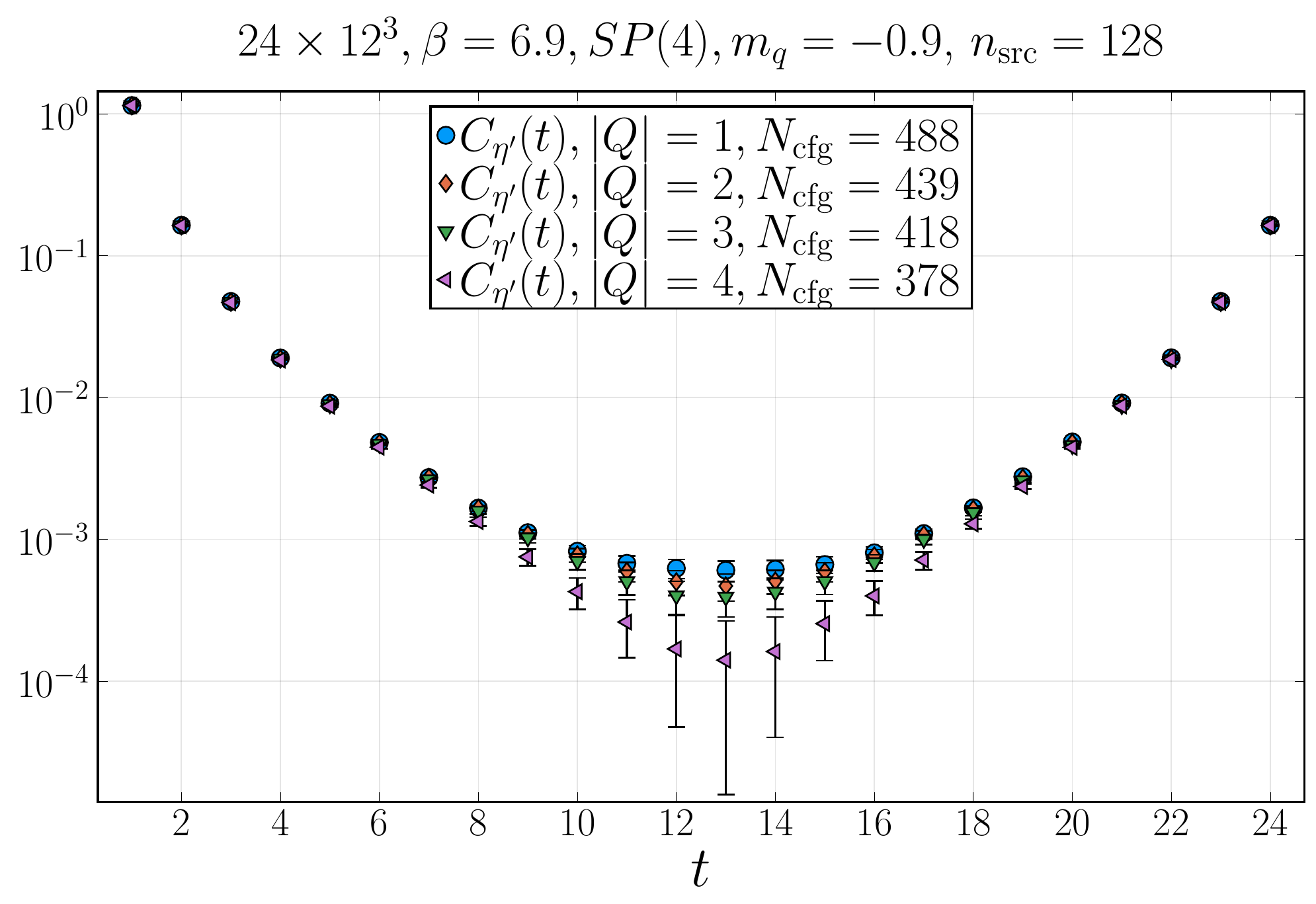}
    \caption{(left) Correlator of the pseudoscalar non-singlet $\pi$ and the pseudoscalar singlet $\eta'$. For visual clarity we multiplied the $\pi$ correlator by a constant factor of $2$. At large times the singlet correlator shows a constant term, while this is absent for the non-singlet case. (right) Correlators of the pseudoscalar singlet for fixed values of the topological charge $Q$. The constant term shows signs of a dependence on $Q$. While the constant is not significantly different for any two examples shown here, the constant appears to be increasing with $|Q|$.}
    \label{fig:eta_example_constant}
\end{figure}

In Sect.~\ref{ssec:op} we noted the occurrence of constant terms in the propagators of both the pseudoscalar singlet $\eta'$ meson and the scalar singlet $\sigma$ meson. This makes it difficult to determine when the excited states in the meson correlator are sufficiently suppressed and a fit can be performed. 
As shown in Eqs.~\eqref{eq:contraction_diagrams} and~\eqref{eq:correlator_constant}, we can circumvent this issue either by direct calculation of $\langle 0 | O | 0 \rangle $, or by performing a numerical derivative. Once we determine the interval $[t_i, t_f]$ where only the ground state contributes, we can also fit the correlator to an exponentially decaying term plus a constant. In Fig.~\ref{fig:eta_example_constant} we give an example of the correlator for the flavor singlet, $\eta'$, and the flavored mesons, $\pi$. The flavor-singlet correlator shows a constant term at large Euclidean times while such a contribution is absent for the $\pi$ meson. This is expected to occur for the disconnected pieces in a finite volume and at finite statistics if the topological sampling is insufficient \cite{Aoki:2007ka, Dimopoulos:2018xkm}. Then, the constant takes the form
\begin{align}
   \label{eq:eta_const} 
   \mathrm{const} \propto \frac{1}{V} \left( \frac{Q^2}{V} - \chi_t - \frac{c_4}{2\chi_t V} \right) + \mathcal O(V^{-3}) + \mathcal O(e^{-m_{\pi} |x| }), 
\end{align}
where $V$ denotes the spatial volume of the lattice, $\chi_t$ is the topological susceptibility and $c_4$ is a coefficient from a saddle-point expansion.
Algorithms that provide an ergodic exploration of topological sectors in Yang-Mills theories and are compatible with our prescription for the boundary conditions have recently been introduced (see, e.g.,~\cite{Bonanno:2020hht,Cossu:2021bgn,Bonanno:2022yjr}). However, they are generally computationally costly, and their adaptation to our model is outside the scope of this work. Hence, in our study, we explore other strategies, based on analyses that account for topological freezing.   
Therefore, we tried to test the relation in Eq.~(\ref{eq:eta_const}) by taking our ensemble with the largest statistics (corresponding to the bare parameters $m_0=-0.90$, $\beta=6.9$ on a $24 \times 12^3$ lattice), measuring the topological charge $Q$ using the same approach as in Ref.~\cite{Bennett:2019jzz}, and smoothening the gauge fields using the gradient flow. We then partition our full statistics into sets of configurations with equal topological charge\footnote{In practice, the topological charge is not strictly an integer on a finite lattice in the employed approach as described for instance in \cite{Bennett:2019jzz} and references therein. We thus round $Q$ to the closest integer.} and compute the correlator for the pseudoscalar singlet, $\eta'$, at fixed $Q$. We depict examples of the correlators for some values of $Q$ with sufficient statistics in Fig.~\ref{fig:eta_example_constant}. The constant term arising is never statistically different for any pair of $Q$'s present in this ensemble. However, we see a slight trend towards a larger constant for larger $|Q|$ as expected from Eq.~\eqref{eq:eta_const}.

In order to test the robustness of our subtraction choice, we report here the mass of the pseudoscalar singlet $\eta'$ for various techniques. We remind the reader that the results reported in Sec.~\ref{sec:results} are based on correlators where the connected part is modelled by a single sum of exponentials as in Eq.~\eqref{eq:correlator_constant}, taking lattice periodicity into account and removing the constant by a numerical derivative. In Tab.~\ref{tab:eta_different_techniques} we compare this to four alternative methods:\footnote{We also applied an entirely different method designed for situations with large statistical noise \cite{Jenny:2022atm}. Also, this approach gave  results consistent with those shown in the main part of the paper for both singlets.} (i) Direct calculation and subtraction of $\langle 0 | O_{\eta'} | 0 \rangle$, (ii) Ignoring the constant and restricting the fit to early time-slices, (iii) Performing a three-parameter fit of the decaying exponential plus a term modelling the constant,\footnote{This procedure gives the numerical value of the constant as a byproduct. We subtract the constant from the correlator and \emph{a posteriori} check that the resulting effective mass shows a plateau.} (iv) Removing the constant using a numerical derivative but without any modelling of the connected part.

Whenever we obtain a signal without an explicit modelling of the connected pieces our results agree within errors. The removal of excited state contamination (as used in methods (i), (ii) and (iii)) leads to masses that are generally slightly lighter. The same pattern has been observed in $\SU{3}$ \cite{Dimopoulos:2018xkm}. We note that the removal of excited state contaminations should not be confused with the removal of the constant contribution to the correlator, as discussed earlier. The explicit calculation of the constant $\langle 0 | O_{\eta'} | 0 \rangle$ in Eq.~\eqref{eq:correlator_constant} does not quantitatively capture the constant in the correlator. The results are almost indistinguishable from not taking the constant into account. For some ensembles (e.g.\ $\Sp{4}$ with $\beta=7.2$) these methods appear to underestimate the meson mass. This is a result of combining the modelling of the connected piece with an insufficient subtraction of the constant. Due to the absence of connected excited states in the correlator, the effective mass is increasing at small $t$, while for large $t$ the constant leads to a decrease of the effective masses. This can lead to the formation of an apparent plateau in the effective mass and thus to a possible underestimation of the meson mass. Overall, we conclude that methods (ii) and (iii) do not appear sufficiently reliable. Modelling the constant as an additional fit parameter did not lead to any significant improvements. In most cases we cannot extract a reliable signal. In the few cases where this is possible the constant term is quantitatively small and this method agrees with the others while providing no improvement at the cost of an additional fit parameter. 

We conclude that the method used throughout the main part of this work has proven to be the most reliable approach among the options considered here. Its results are always consistent with forgoing the explicit removal of subtracted states, and the removal of the additional constant through taking the derivative avoids any further estimations of the topological constant terms at the expense of a shorter plateau in the effective masses  and thus, a smaller interval for fitting the correlator.

We find a  different behavior for the scalar singlet. The constant term is not related to an insufficient sampling of all topological sectors but arises due to the vacuum quantum numbers of the scalar singlet. In addition, the modelling of the connected pieces is less important, since the non-singlet state appears generally heavier than the singlet states and the connected pieces show a stronger exponential decay. In this case the direct estimation of the constant term $\langle 0 | O_{\sigma} | 0 \rangle$ in Eq.~\eqref{eq:correlator_constant} appears to be quantitatively reliable. Still, in some cases the modelling of the connected pieces can extend the plateau in the effective mass to lower timeslices $t$. Since the constant is several order of magnitude larger than the actual signal of the $\sigma$ state, a direct modelling of it as a fit parameter is infeasible and the constant can also not be ignored in the analysis. In Tab.~\ref{tab:sigma_different_techniques} we compare the approach used in the main part of this paper to: (i) both a numerical derivative and a direct calculation and subsequent subtraction of the vacuum term $\langle 0 | O_{\sigma} | 0 \rangle$, (ii) only direct subtraction of the vacuum terms as in \cite{Kunihiro:2003yj}, (iii) a numerical derivative without an explicit subtraction of excited states in the connected pieces and without a direct subtraction of $\langle 0 | O_{\sigma} | 0 \rangle$.
\begin{table}
    \begin{tabular}{|c|c|c|c|c|c|c|c|c|c|}
	\hline
	& $\beta$ & $m_0$ & $L$ & $T$ & $m_{\eta'}$ & $m_{\eta'}^{\rm (i)}$ & $m_{\eta'}^{\rm (ii)}$ & $m_{\eta'}^{\rm (iii)}$ & $m_{\eta'}^{\rm (iv)}$\\
	\hline\hline
	SU(2)&2.0&-0.947&20&32&-&-&-&-&-\\
	SU(2)&2.0&-0.94&14&24&0.67(6)&0.699(14)&0.572(14)&-&0.67(6)\\
	SU(2)&2.0&-0.935&16&32&0.60(3)&0.67(3)&0.61(5)&0.60(3)&-\\
	SU(2)&2.0&-0.93&14&24&0.65(3)&-&-&-&0.68(5)\\
	SU(2)&2.0&-0.925&14&24&0.634(16)&-&-&-&0.63(8)\\
	SU(2)&2.0&-0.92&12&24&0.665(9)&-&-&-&0.66(3)\\
	SU(2)&2.0&-0.9&12&24&0.770(16)&-&-&-&0.79(8)\\
	SU(2)&2.0&-0.88&10&20&0.842(5)&0.855(14)&0.855(14)&-&-\\
	\hline\hline
	Sp(4)&7.2&-0.799&32&40&-&0.37(2)&0.37(2)&-&0.57(6)\\
	Sp(4)&7.2&-0.794&28&36&0.397(16)&0.368(12)&0.368(12)&-&0.47(7)\\
	Sp(4)&7.2&-0.79&24&36&0.387(13)&-&-&-&0.36(6)\\
	Sp(4)&7.2&-0.78&24&36&0.418(7)&0.43(2)&0.45(2)&-&0.43(5)\\
	Sp(4)&7.2&-0.77&24&36&0.456(8)&0.450(6)&0.450(6)&0.459(7)&-\\
	Sp(4)&7.2&-0.76&16&36&-&0.511(13)&0.512(13)&-&0.59(3)\\
	\hline\hline
	Sp(4)&6.9&-0.924&24&32&-&-&-&-&0.60(8)\\
	Sp(4)&6.9&-0.92&24&32&-&0.51(4)&0.52(4)&0.486(16)&0.40(6)\\
	Sp(4)&6.9&-0.92&16&32&0.49(3)&0.46(2)&0.46(2)&0.50(2)&0.45(13)\\
	Sp(4)&6.9&-0.91&16&32&0.560(14)&0.59(4)&0.59(4)&0.560(13)&0.59(4)\\
	Sp(4)&6.9&-0.91&14&24&0.541(9)&0.58(3)&0.58(3)&-&-\\
	Sp(4)&6.9&-0.9&16&32&0.611(9)&-&-&-&0.63(3)\\
	Sp(4)&6.9&-0.9&14&24&0.619(16)&0.614(12)&0.615(12)&0.620(9)&0.63(3)\\
	Sp(4)&6.9&-0.9&12&24&0.610(6)&0.620(15)&0.620(15)&0.612(5)&-\\
	Sp(4)&6.9&-0.89&14&24&0.69(2)&0.680(16)&0.681(16)&0.69(2)&0.72(4)\\
	Sp(4)&6.9&-0.89&12&24&0.661(9)&0.660(5)&0.660(5)&0.660(10)&-\\
	Sp(4)&6.9&-0.87&12&24&0.782(13)&0.80(4)&0.80(4)&-&0.80(2)\\
	Sp(4)&6.9&-0.87&10&20&0.764(9)&0.763(6)&0.763(6)&-&-\\
	\hline\hline
\end{tabular}
    \caption{Determination of the masses of the pseudoscalar singlets using different techniques for removing the constant term in the correlator. We compare the method used in the main part of this work to: (i) Direct calculation and subtraction of $\langle 0 | O_{\eta'} | 0 \rangle$; (ii) Ignoring the constant and restricting the fit to early time-slices; (iii) Performing a three-parameter fit of the decaying exponential plus a term modelling the constant; and (iv) Removing the constant using a numerical derivative but without any modelling of the connected part.}
    \label{tab:eta_different_techniques}
\end{table}

\begin{table}
    \begin{tabular}{|c|c|c|c|c|c|c|c|c|}
	\hline
	& $\beta$ & $m_0$ & $L$  & $T$  & $m_{\sigma}$ & $m_{\sigma}^{(i)}$ & $m_{\sigma}^{\rm (ii)}$ & $m_{\sigma}^{\rm (iii)}$\\
	\hline\hline
	SU(2)&2.0&-0.947&20&32&0.53(4)&0.53(3)&0.58(5)&0.54(4)\\
	SU(2)&2.0&-0.94&14&24&-&0.64(5)&0.61(4)&0.64(5)\\
	SU(2)&2.0&-0.935&16&32&-&0.47(10)&0.55(8)&0.47(10)\\
	SU(2)&2.0&-0.93&14&24&-&-&0.62(9)&-\\
	SU(2)&2.0&-0.925&14&24&-&-&-&-\\
	SU(2)&2.0&-0.92&12&24&-&0.71(7)&0.75(13)&0.72(7)\\
	SU(2)&2.0&-0.9&12&24&-&-&-&-\\
	SU(2)&2.0&-0.88&10&20&-&-&-&-\\
	\hline\hline
	Sp(4)&7.2&-0.799&32&40&0.36(5)&0.35(8)&0.41(4)&0.38(8)\\
	Sp(4)&7.2&-0.794&28&36&-&0.55(8)&-&0.55(8)\\
	Sp(4)&7.2&-0.79&24&36&0.56(6)&0.56(6)&0.48(7)&0.65(12)\\
	Sp(4)&7.2&-0.78&24&36&-&-&0.55(6)&-\\
	Sp(4)&7.2&-0.77&24&36&-&-&-&-\\
	Sp(4)&7.2&-0.76&16&36&0.64(12)&0.64(7)&-&-\\
	\hline\hline
	Sp(4)&6.9&-0.924&24&32&0.46(3)&0.46(3)&0.45(3)&0.48(7)\\
	Sp(4)&6.9&-0.92&24&32&0.42(2)&0.43(3)&0.45(3)&-\\
	Sp(4)&6.9&-0.92&16&32&0.45(6)&0.40(8)&0.42(7)&0.37(11)\\
	Sp(4)&6.9&-0.91&16&32&-&-&-&0.71(8)\\
	Sp(4)&6.9&-0.91&14&24&0.41(3)&0.41(3)&-&-\\
	Sp(4)&6.9&-0.9&16&32&-&-&-&-\\
	Sp(4)&6.9&-0.9&14&24&0.57(4)&0.56(4)&0.48(5)&0.51(10)\\
	Sp(4)&6.9&-0.9&12&24&0.55(2)&0.56(4)&-&0.57(3)\\
	Sp(4)&6.9&-0.89&14&24&0.57(9)&0.56(9)&0.61(9)&0.59(9)\\
	Sp(4)&6.9&-0.89&12&24&0.62(7)&-&0.64(4)&-\\
	Sp(4)&6.9&-0.87&12&24&0.80(15)&0.76(11)&-&0.72(7)\\
	Sp(4)&6.9&-0.87&10&20&-&0.70(6)&-&-\\
	\hline\hline
\end{tabular}
    \caption{Results for the masses of the scalar singlet $\sigma$ using our standard approach of a numerical derivative as well as (i) both a numerical derivative and a direct calculation of the vacuum term $\langle 0 | O_{\sigma} | 0 \rangle$, (ii) only direct calculation of the vacuum term, and (iii) a numerical derivative without an explicit subtraction of excited states in the connected piece when possible.}
    \label{tab:sigma_different_techniques}
\end{table}

\section{Comparison between excited state subtraction and smeared connected diagrams}
\label{sec:smeared_connected}

In Sect.~\ref{ssec:signalboost} we noted that the signal of the singlet mesons can be extended to smaller time separations, $t$, if we replace its connected contribution by approximating it with a single exponential term having the energy of the non-singlet meson. This removes all the excited state contaminations of the connected piece. 

A similar effect can be obtained by using smearing techniques on the connected piece. This can increase the overlap of the source operator with the ground state of the non-singlet and reduce the contribution of excited states. Recently, this approach has been implemented, tested, and shown to work in~$\Sp{4}$ gauge theories~\cite{Hsiao:2022kxf}. These developments allow us to compare our excited-state subtraction technique.\footnote{We thank the authors of \cite{Hsiao:2022kxf,Lee:2022elf} for performing smeared measurements on a set of our configurations for comparison prior to publication.}

\begin{figure}
    \centering
    \includegraphics[width=.8\textwidth]{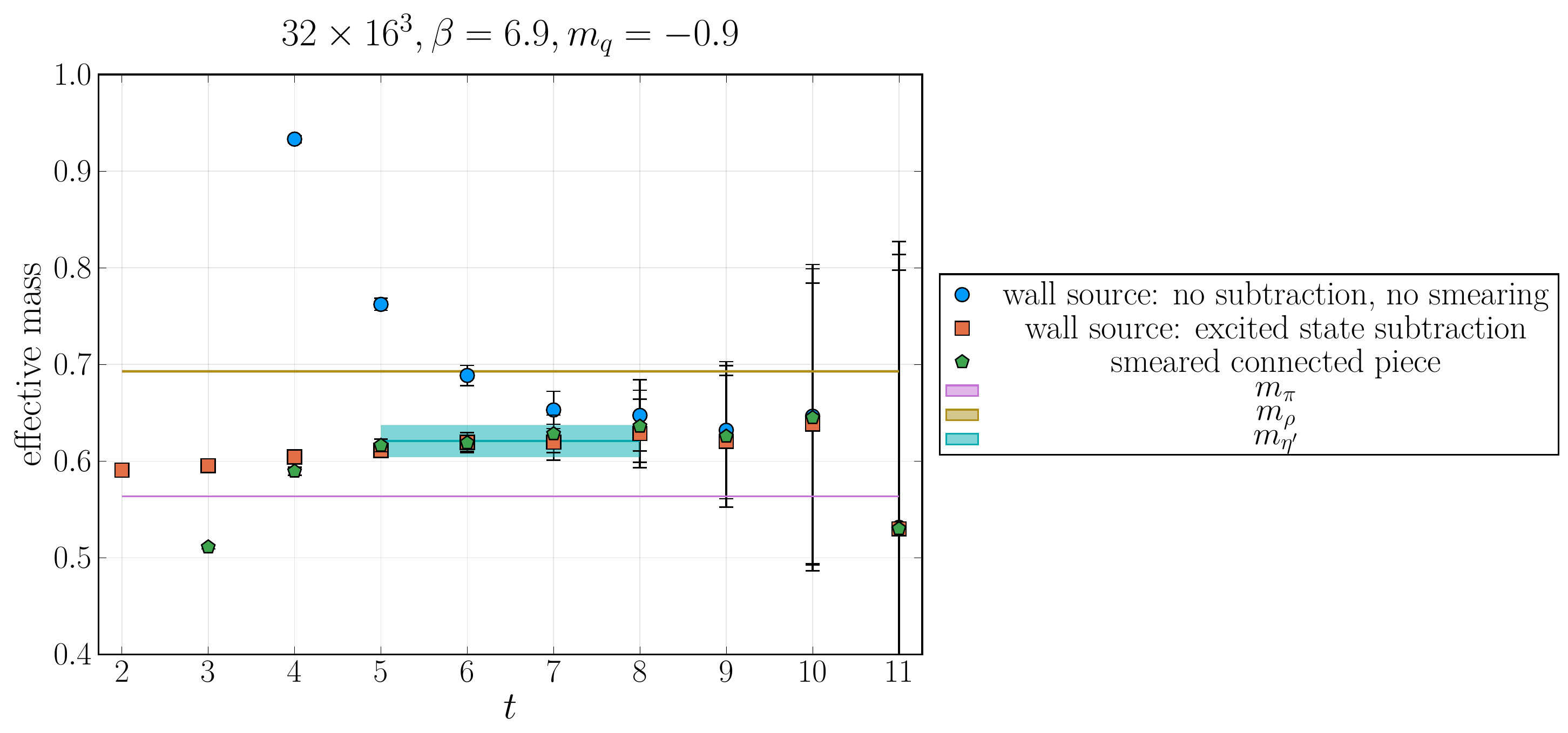}
    \caption{Comparison between excited state subtraction (orange squares), obtained by modelling the connected part of the correlator as a single exponential term, with  smeared operators (green pentagon), on a preliminary set of configurations, for a single ensemble. For reference, we also plot the correlator without excited state subtraction and without smearing (blue circles). 
    }
    \label{fig:smeared_connected}
\end{figure}

In order to compare the two techniques we need to apply smearing to only the connected piece. However, the use of smearing techniques leads to an overall change of normalization. Applying Wuppertal smearing~\cite{Gusken:1989ad} with $N_1$ steps at the source and $N_2$ steps at the sink leads to an asymptotic correlator of the form
\begin{align}
    C_{N_1,N_2}(t \to \infty) = \alpha_{N_1} \alpha_{N_2} e^{-m_{\rm conn}t}\,,
\end{align}
where the normalization of unsmeared point sources is recovered for the choice of the parameters $\alpha_{N_1} = \alpha_{N_2} = \alpha_{0}$. We consider  two sets of correlators with the smearing steps $(N_1, N_2)=(N,0)$ and $(N_1, N_2)=(N,N)$, to restore the normalization as the point source. We define a new correlator
\begin{align}
    \label{eq:normalized_smeared_correlator}
    C^{\rm smeared}_{\rm conn.}(t) \equiv \frac{C_{N,0}(t)^2}{C_{N,N}(t)}\,,
\end{align}
by squaring the connected correlator with $N$ steps of source smearing and no sink smearing and divide that by the connected correlator with an equal amount of smearing steps at both the source and the sink. This correlator has the same large-$t$ behavior and the same normalization as a non-smeared one. From this we then construct the full correlator of the singlet meson. In Fig.~\ref{fig:smeared_connected} we compare the full singlet correlator obtained from Eq.~\eqref{eq:normalized_smeared_correlator} using Wuppertal smearing with $N=60$ smearing steps, to the correlator obtained using the single-exponential modelling and subtraction of the connected piece. We see that the  subtracted correlator and the smeared correlator agree remarkably well in the interesting, plateau region.

\bibliography{apssamp}

\end{document}